\documentclass[twocolumn]{aastex631}

\usepackage{xcolor}
\usepackage{rotating}
\usepackage{amsmath}
\usepackage{verbatim}
\usepackage{csquotes}

\shorttitle{Core Mass Function in the Galactic Center}
\shortauthors{Kinman et al.}

\newcommand{\msol}{$M_{\odot}$}
\newcommand{\as}{$^{\prime\prime}$}

\defcitealias{cheng2018}{I}
\defcitealias{liu2018}{II}
\defcitealias{oneill2021}{III}

\begin{document}

\title{The Core Mass Function Across Galactic Environments. IV.\\The Galactic Center}

\author[0009-0008-6570-9287]{Alva V. I. Kinman}
\affiliation{Dept. of Space, Earth \& Environment, Chalmers University of Technology, Gothenburg, Sweden}

\author[0000-0002-6362-8159]{Maya A. Petkova}
\affiliation{Dept. of Space, Earth \& Environment, Chalmers University of Technology, Gothenburg, Sweden}

\author[0000-0002-3389-9142]{Jonathan C. Tan}
\affiliation{Dept. of Space, Earth \& Environment, Chalmers University of Technology, Gothenburg, Sweden}
\affiliation{Dept. of Astronomy, University of Virginia, Charlottesville, Virginia 22904, USA}

\author[0000-0001-5551-9502]{Giuliana Cosentino}
\affiliation{Dept. of Space, Earth \& Environment, Chalmers University of Technology, Gothenburg, Sweden}
\affiliation{European Southern Observatory, Karl-Schwarzschild-Strasse 2, D-85748 Garching, Germany}

\author[0000-0002-8691-4588]{Yu Cheng}
\affiliation{National Astronomical Observatory of Japan, Mitaka, Tokyo, 181-8588, Japan}

\begin{abstract}
The origin of the stellar Initial Mass Function (IMF) and how it may vary with galactic environment is a matter of debate. Certain star formation theories involve a close connection between the IMF and the Core Mass Function (CMF) so it is important to measure this CMF in a range of Milky Way locations. Here we study the CMF of three Galactic Center clouds: G0.253+0.016 (\enquote{The Brick}), Sgr~B2 (Deep South field) and Sgr~C. We use ALMA 1~mm continuum images and identify cores as peaks in thermal dust emission via the dendrogram algorithm. We develop a completeness correction method via synthetic core insertion, utilizing a realistic mass-dependent size distribution. A power law $\text{d}N/\text{d}\log M \propto M^{-\alpha}$ is fit to the CMFs $>2\:M_\odot$. 
The Brick has a Salpeter-like index $\alpha=1.28\pm0.09$, while the other regions have shallower indices: Sgr~C has $\alpha=0.99\pm0.06$; Sgr~B2-DS has $\alpha=0.70\pm0.03$. When smoothed to a common resolution, the differences between the Brick and the others increase as we obtain $\alpha=1.36\pm0.12$, $\alpha=0.66\pm0.06$ and $\alpha=0.62\pm0.04$, respectively, for masses $\gtrsim3\:M_\odot$. 
Furthermore, we analyze the spatial distribution and mass segregation of cores: Sgr~C and Sgr~B2-DS show signs of mass segregation, but the Brick does not. We compare our results to several other CMFs from different Galactic regions derived with the same methods. Finally, we discuss how our results may help define an evolutionary sequence of star cluster formation and be used to test star formation theories.
\end{abstract}

 
\section{Introduction} \label{S:intro}

The stellar initial mass function (IMF) is fundamental for many areas of astrophysics, from star formation to galaxy evolution. In particular, any time an unresolved stellar population is studied, the IMF is needed to convert from luminosity to quantities such as stellar mass and star formation rate. The IMF is also crucial for models of how a given stellar population produces radiative, mechanical and chemical enrichment feedback on its surrounding interstellar medium and galaxy.
However, the origin of the IMF is still a matter of debate.

The high-mass end of the IMF approximately follows a power law of the form
\begin{equation}  
\frac{d N}{d\: {\rm log}\: M}\propto{M}^{-\alpha}.
\label{eq:powerlaw}
\end{equation}
The power law index was found by \cite{salpeter1955} to be $\alpha=1.35\pm 0.2$, and this value still holds in some recent studies. The IMF has been found to have a peak at $\sim 0.2$ \msol \citep[e.g., see][for a review]{offner2014}.

The IMF appears to be universal in a wide range of environments \citep{bastian2010,offner2014}, although some variation has been claimed in extreme environments such as massive starburst regions \citep[e.g.,][]{schneider2018} and high redshift galaxies \citep[e.g.,][]{vanDokkum2017}. A top-heavy ($\alpha=0.80\pm0.1$) IMF has also been claimed in the Arches Cluster in the Galactic Center region by \citet{hosek2019}, although the result assumes that dynamical processes, such as tidal stripping and mass segregation, have not altered the IMF compared to the mass function that is observed in the current epoch.

In order to understand the origin of the IMF, it is important to study early phases of star formation, including prestellar and protostellar cores. The mass distribution of the cores, i.e., the core mass function (CMF), can be measured and by comparison with the IMF, various star formation theories can be tested. The ``Core Accretion'' class of theoretical models \citep[e.g.,][]{1987ARA&A..25...23S,1997ApJ...476..750M,mckee2003} assume dense molecular gas clumps fragment into populations of prestellar cores. Then, in a protostellar core phase, these objects form a single, central, rotationally-supported disk that feeds a single protostar or a small-$N$ multiple originating via disk fragmentation. A significant fraction, $\epsilon_{\rm core}\sim 0.5$, of the initial mass of the core is expected to end up in the primary star that forms from the core \citep[e.g.,][]{2000ApJ...545..364M}, although this may vary with mass, pressure, and metallicity of the environment \citep[e.g.,][]{2017ApJ...835...32T,2018ApJ...861...68T}. In addition, if the collapse of the core is relatively fast, i.e., on a timescale similar to its local free-fall time, then there is relatively little time for significant amounts of additional gas to join the core in the protostellar phase from the lower-density surrounding clump. For example, in the fiducial Turbulent Core Accretion (TCA) model of \citet{mckee2003}, a core is expected to interact with surrounding clump gas at a rate that is similar to the mass accretion rate to the protostar. However, as discussed by \citet{mckee2003}, given the dynamical influence of magnetic fields, it appears unlikely that this gas becomes bound to the core.

A number of theoretical proposals for the origin of the prestellar CMF have been proposed. One class of such models invokes the gas conditions produced in shock compressed regions arising in supersonically turbulent molecular clouds with relatively weak magnetic fields \citep[e.g.,][]{2002ApJ...576..870P,2008ApJ...684..395H}. Alternative models involve core formation regulated by ambipolar diffusion in more strongly magnetized clouds \citep[e.g.,][]{2009MNRAS.399L..94K}. Thus it is possible for models of the prestellar CMF to have environmental dependence. For example, more massive cores may require denser clump conditions if they form via a bottom-up coagulation process from lower-mass prestellar cores. Or they may require more strongly magnetized conditions to prevent their fragmentation. Observational constraints on the degree of magnetization of star-forming molecular clouds are relatively challenging to obtain and there are significant uncertainties on the relative importance of turbulence, magnetic fields, and self-gravity \citep[e.g.,][]{2012ARA&A..50...29C}. Some more recent studies have concluded that magnetic fields are dynamically important and that turbulence is present at a trans- or sub-Alfv\'enic level \citep[e.g.,][]{2015ApJ...799...74P,2015Natur.520..518L,2024arXiv240111560L}.

However, as discussed, even if a prestellar CMF theory is agreed upon, a model of star formation efficiency from a given core is still needed to connect to the stellar IMF \citep[e.g.,][]{1996ApJ...464..256A,2000ApJ...545..364M,2017ApJ...835...32T}. In the simple limiting case of constant $\epsilon_{\rm core}$ and negligible core mass gain during the protostellar phase, one would expect the prestellar core mass function to have a similar form as the stellar IMF. The protostellar core mass function is typically estimated via measurements of core envelope masses that exclude the mass of the protostar itself. Its connection to the IMF is less straightforward, even in this limiting case, since one is sampling over the evolutionary phase of the protostars.

In contrast to Core Accretion, ``Competitive Accretion'' models \citep[e.g.,][]{bonnell2001,2022MNRAS.512..216G} involve most of the mass of the star being accreted from the clump after the prestellar core phase has ended, i.e., during the protostellar phase. The fraction of clump-fed mass is expected to be largest for the more massive stars, but is potentially even significant for $\sim1\:M_\odot$ cases. Indeed competitive Bondi-Hoyle accretion has been involved as the process that creates the high-mass power law tail of the stellar IMF \citep[e.g.,][]{2005MNRAS.356.1201B,2022MNRAS.515.4929G}. However, since the prestellar core mass and the instantaneous protostellar envelope mass have little direct connection to the final stellar mass, one does not expect a close correspondence between these CMFs and the IMF. In terms of environmental dependence, in general, one expects competitive accretion to be more important in regions of high protostellar density and to naturally produce primordial mass segregation of the protostellar population, i.e., with massive stars forming at central locations in a protocluster. However, one should note that Core Accretion models, such a bottom-up coagulation model, may also produce such primordial mass segregation in prestellar core populations. This is because coagulation of pre-stellar cores is expected to happen at higher rates in denser environments.

In the solar neighborhood, the CMF seems to closely resemble the IMF, although scaled up in mass by a factor of about three \citep{motte1998, testi1998, andre2010, konyves2015}. In more distant regions, there have been indications of a CMF power law index that is more top-heavy compared to the Salpeter value \citep[e.g.,][]{motte2018,kong2019,sanhueza2019}. In addition, \citet{lu2020} have reported a top-heavy CMF in four massive clouds in the Central Molecular Zone (CMZ), including Sgr~C.

Recently, the ALMA-IMF program has investigated the CMF in the massive star-forming complex W43 \citep{pouteau2022,pouteau2023,nony2023}. \citet{pouteau2022} studied the CMF of the star-forming MM2-MM3 ridge and found it to have an index of $\alpha=0.93^{+0.07}_{-0.10}$ based on a sample of $\sim 200$ cores. \cite{pouteau2023} further investigated the dependence of the CMF on the evolutionary stage, 
finding the CMF to vary from Salpeter-like in quiescent regions to significantly top-heavy in regions undergoing active star formation. Additionally, \citet{nony2023} divided the total core sample into prestellar and protostellar cores. The prestellar cores were found to have a Salpeter-like CMF slope, meaning that the top-heavy shape of the total CMF was caused by the protostellar cores. The ALMA-IMF program has also reported similar findings in the W33 region \citep{armante2024}, where the prestellar CMF was found to be relatively bottom-heavy while the protostellar CMF was top-heavy.

In Paper I in our series, \cite{cheng2018} measured the CMF using the dendrogram algorithm \citep{rosolowsky2008} to select cores as 1.3~mm emission peaks in the massive protocluster G286.21+17. Completeness corrections for flux and number recovery were estimated, resulting in a final derived value of $\alpha=1.24\pm 0.17$ for cores $\gtrsim 1\:M_\odot$. In Paper II, \cite{liu2018} studied the CMF with the same methods in a sample of 30 clumps selected from seven Infrared Dark Clouds (IRDCs), and found a shallower index of $\alpha=0.86\pm0.11$.
In Paper III, \cite{oneill2021} measured the CMF in 28 massive protoclusters using data from the ALMAGAL survey, finding $\alpha= 0.98\pm0.08$ when fitting masses $\gtrsim 5\:M_\odot$.

In this paper, we study the CMF of three regions in the CMZ of the Galactic Center: G0.253+0.016 (a.k.a. \enquote{The Brick}), the deep south region of Sagittarius B2 (Sgr~B2-DS) and Sagittarius C (Sgr~C). The CMZ, i.e., the region within a radius of $\sim 300$ pc from the Galactic center \citep[e.g.,][]{henshaw2023}, contains a collection of molecular clouds. The region is known for its extreme physical conditions: gas number densities $n_{\rm H}>10^4\:{\rm cm}^{-3}$ \citep[e.g.,][]{paglione1998,jones2013}, magnetic field strengths of $\sim1$ mG \citep{pillai2015,federrath2016,mangilli2019}, turbulent Mach numbers $\mathcal{M}\sim 30$ \citep{rathborne2014} and gas temperatures of $>60\:$ K \citep[e.g.][]{ginsburg2016}.  
It is therefore an example of a relatively extreme star-forming environment that allows more stringent testing of predictions of star formation theories.

The Brick has a mass of $\sim10^5\:M_\odot$ and a radius of about 3~pc \citep{lis1994,longmore2012}, making it one of the most massive high-density molecular clouds in the Galaxy \citep[see, e.g.,][]{tan2014}.
However, apart from a water maser \citep{lis1994}, where a small forming stellar cluster has been found \citep{walker2021}, the Brick 
only shows modest signs of star formation \citep[e.g.,][]{immer2012,mills2015}.
On the other hand, the cloud complex Sgr~B2 is one of the most active star-forming regions in the Galaxy. The most intense bursts of star formation are occurring in the Sgr~B2-N and Sgr~B2-M dense clumps, but significant star formation activity has also been detected away from these sites \citep{ginsburg2018}.
Sgr~C is also a known star formation site, including evidence for massive protostellar sources \citep[e.g.,][]{lu2019}.

In this paper, we mostly follow the same methods as the previous papers in this series. However, we develop a new method of completeness correction via artificial core insertion that takes the finite size of cores into account. In \S\ref{S:observations} we describe the observational data. In \S\ref{S:analysis} we present our analysis methods, including core identification, mass estimation and the new flux and number corrections. In \S\ref{S:results} we present the derived CMFs and power law fits. We also compare CMFs derived by the old and new correction method, including CMFs from Papers \citetalias{cheng2018}-\citetalias{oneill2021}. Furthermore, we investigate the spatial clustering and mass segregation of the cores in the CMZ regions. 
We discuss the implications of our results and possible sources of uncertainties in \S\ref{S:discussion}, and finally present our conclusions in \S\ref{S:conclusion}.

\section{Observations} \label{S:observations}

\subsection{Observational Data}
\label{met:observations}

We utilize continuum images from ALMA science archive obtained by the 12~m array of the ALMA telescope in band 6: in particular, mosaics of the three CMZ regions G0.253+0.016 (the Brick), Sagittarius B2-Deep South (Sgr~B2-DS) and Sagittarius C (Sgr~C). They were obtained from the ALMA archive with IDs 2012.1.00133.S (PI: G. Garay), 2017.1.00114.S (PI: A. Ginsburg) and 2016.1.00243.S (PI: Q. Zhang), respectively. The Sgr~B2-DS and Sgr~C datasets have previously been used to examine core properties \citep{lu2020,lu2021,meng2022,jeff2024}. We have used the archival images directly, without reimaging. The Brick image was however regridded to have a similar number of pixels per beam as the other two images, using the \textit{imregrid} task in CASA 6.5.3.

The Brick was observed by 32 antennas in Cycle 1, with baselines in the range 15-360~m.  Four spectral windows with bandwidth 1.875 GHz and center frequencies 251.521, 250.221, 265.523 and 267.580 GHz were used. The resulting continuum image has a central frequency of 258 GHz, corresponding to a wavelength of 1.16~mm. Calibration was done with the sources J1924-2914, J2230-1325, J1744-3116 and Neptune. The mosaic consists of 140 pointings.

Sgr~B2 was observed by 43 antennas in Cycle 5, with baselines in the range 15-783~m.  Four spectral windows with bandwidth 1.875 GHz and center frequencies 217.366, 219.199, 231.308 and 233.183 GHz were used. The resulting continuum image has a central frequency of 225 GHz, corresponding to a wavelength of 1.33~mm. Calibration was done with the sources J1924-2914 and J1744-3116. The mosaic consists of 11 pointings.

Finally, Sgr~C was observed by 39 antennas in Cycle 4, with baselines in the range 15-460~m.  Four spectral windows with bandwidth 1.875~GHz and center frequencies 217.915, 219.998, 232.110 and 234.110 GHz were used. The resulting continuum image has a central frequency of 226~GHz, corresponding to a wavelength of 1.33~mm. Calibration was done with the sources J1924-2914, J1742-1517 and J1744-3116. The mosaic consists of 9 pointings.

The synthesized FWHM beam sizes of the images are $1.03$\as $\times 0.855$\as (Position angle=$78^\circ$) for the Brick, $0.46$\as $\times 0.37$\as (PA=$-86^\circ$) for Sgr~B2 and $0.80$\as $ \times 0.60$\as (PA=$84^\circ$) for Sgr~C. Maximum recoverable scales are $10.6''$, $6.5''$ and $6.4''$ for the Brick, Sgr~B2 and Sgr~C, respectively.

Additionally, we used 1.1 mm continuum images from the Bolocam Galactic Plane Survey \citep{ginsburg2013_BOLOCAM,bolocam_data}, to estimate the total mass within the ALMA fields of view. These images have a beam FWHM of 33\as.

\subsection{Noise}

To estimate the RMS noise of the non-primary-beam-corrected ALMA images, beam-sized patches were randomly placed in the image, and the mean intensity in each patch was obtained. If the mean was larger than 0.1 times the maximum signal in the image, the patch was considered as containing signal and discarded. The final result was found to be insensitive to this threshold. This was repeated 10,000 times. A Gaussian distribution was then fit to the distribution of intensities. The RMS noise $\sigma$ was taken to be the standard deviation of the Gaussian. To lessen the effect of random sampling, the above was repeated 100 times and the median of $\sigma$ was used.
The obtained noise levels were 0.174 mJy~beam$^{-1}$ for the Brick, 0.111 mJy~beam$^{-1}$ for Sgr~B2 and 0.127 mJy~beam$^{-1}$ for Sgr~C.

\section{Analysis Methods} \label{S:analysis}

\subsection{Core Identification}
\label{met:core_identification}

Following Papers \citetalias{cheng2018}-\citetalias{oneill2021}, cores were identified using the dendrogram algorithm \citep{rosolowsky2008}, implemented in the python package \texttt{astrodendro}\footnote{\url{https://dendrograms.readthedocs.io/en/stable/}}. Dendrogram identifies peaks in data and sorts them in a hierarchical structure, consisting of \enquote{branches} (containing substructure) and \enquote{leaves} (containing no substructure). We used the same dendrogram parameters as previous papers. In particular, structures were required to have a minimum intensity $F_{\rm min}=4\sigma$, and substructures (branches and leaves) to have a minimum intensity increase $\delta=1\sigma$. Finally, structures were required to contain a minimum number of pixels $N_{\rm min}$ corresponding to half the area of the beam, defined by the following equation:
\begin{equation}
    N_{\rm min}=\frac{\pi\theta_{\rm maj}\theta_{\rm min}}{8 A_{\rm pix}},
\end{equation}
where $\theta_{\rm maj}$ and $\theta_{\rm min}$ are the major and minor full width half maxima of the beam in arcseconds and $A_{\rm pix}$ is the area of each pixel in arcseconds squared. Cores were defined as the dendrogram leaves.

As in Paper \citetalias{liu2018} and \citetalias{oneill2021}, cores were identified in the non-primary-beam-corrected images. This was to obtain a uniform noise level across the map. Subsequently, we use the primary-beam-corrected fluxes of the identified cores for constructing the CMF. Core fluxes were computed by summing the flux of every pixel belonging to the dendrogram leaf, without background subtraction \citep[i.e. by the bijection paradigm in][]{rosolowsky2008}. Furthermore, the detection of cores was restricted to the parts of the mosaic where the primary beam response exceeded 0.5.
 
\subsection{Core Mass Estimation}
\label{met:mass_estimation}

Core masses were estimated from thermal dust emission assuming this emission to be optically thin, as in Papers \citetalias{cheng2018,liu2018,oneill2021}. The mass surface density is given by 
\begin{eqnarray}
\Sigma_{\rm mm} & = & 0.369 \frac{F_\nu}{\rm mJy}\frac{(1^{\prime\prime})^2}{\Omega} \frac{\lambda_{1.3}^3}{\kappa_{\nu,0.00638}}
 \nonumber\\
 & & \times  \left[{\rm exp}\left(0.553 T_{d,20}^{-1}
  \lambda_{1.3}^{-1}\right)-1\right]\:{\rm g\:cm^{-2}},
\label{eq:Sigmamm}
\end{eqnarray}
where $F_{\nu}$ is the total integrated flux over solid angle
$\Omega$, $\kappa_{\nu,0.00638}\equiv\kappa_\nu/({\rm
6.38\times10^{-3}\:cm^2\:g}^{-1})$ is the dust absorption
coefficient, $\lambda_{1.3}=\lambda/1.30\:{\rm mm}$ and
$T_{d,20}\equiv T_d/20\:{\rm K}$ with $T_d$ being the dust temperature. 

Since we do not have temperature data for each core, a dust temperature of 20~K has been assumed, as in Paper \citetalias{cheng2018}, \citetalias{liu2018} and \citetalias{oneill2021}. Such a temperature has been found to be a typical of models of lower-mass protostellar cores \citep{zhang2015}. It is also consistent with dust temperature measurements in the CMZ using Herschel \citep{longmore2012,extaluze2013,ginsburg2016,kauffmann2017}. For example, \cite{kauffmann2017} measured dust temperatures of 17-25 K for various CMZ clouds including the Brick and Sgr~C. \cite{extaluze2013} measured $\sim22$ K in Sgr~B2-DS. However, it is possible that the temperature varies among the cores. If a temperature of 15 K was assumed instead, it would increase the calculated mass by a factor of 1.48. If the temperature was increased to 30 K, the mass would be reduced by a factor of 0.604. Note that there may be systematic temperature variations: for example, brighter cores could be warmer. In that case, the more massive cores would have their masses overestimated by our current method.

To obtain the value of the dust absorption coefficient for 1.3 mm emission, we assumed an
opacity per unit dust mass $\kappa_{\rm 1.3mm,d}=0.899\: {\rm cm^2
 \ g}^{-1}$ (based on the moderately-coagulated thin ice mantle model of \citealt{ossenkopf1994}), which results in $\kappa_{\rm 1.3mm}= {\rm
  6.38\times10^{-3}\:cm^2\:g}^{-1}$ using a
gas-to-refractory-component-dust ratio of 141 \citep{draine2011}. 
Since the Brick image has a wavelength of 1.16 mm, the value of $\kappa_{\rm 1.16mm,d}$ was estimated by linear interpolation between $\kappa_{\rm 1.0mm,d}$ and $\kappa_{\rm 1.3mm,d}$ in \cite{ossenkopf1994}.

To obtain core masses, the mean mass surface density is multiplied by the area of the core: 
\begin{equation}  
M  =  \Sigma_{\rm mm} A = 0.113 \frac{\Sigma_{\rm mm}}{\rm g\:cm^{-2}} \frac{\Omega}{(1^{\prime\prime})^2} \left(\frac{d}{\rm 1\:kpc}\right)^2  \:M_\odot,
\label{eq:coremass} 
\end{equation}
where $d$ is the distance to the CMZ. We adopt a distance of 8.3 kpc to all regions, consistent with the distance of 8,277 pc to Sgr~A* found by \citet{gravity2022}. The individual clouds may be displaced along the line of sight on the order of a few hundred pc compared to Sgr~A*, but the precise morphology of the CMZ is not settled \citep[see, e.g.,][]{henshaw2023}. Therefore we do not apply disparate individual distances to the regions. An error of 5\% ($\sim$400 pc) in the estimated distances results in a 10\% error in the masses. However, note that the distance uncertainties would only affect the systematic mass difference for the core population between different clouds, so the CMF index of an individual cloud would not be affected by this uncertainty.

\subsection{Core Flux Recovery and Completeness Corrections}\label{met:recovery}
To obtain the ``true'' CMF, corrections for incompleteness need to be made. Firstly, dendrogram excludes pixels with intensities less than $F_{\rm min}$. This means that some of the flux of the cores is lost. Secondly, the algorithm may miss small and faint cores entirely. To correct for these two effects, a flux recovery fraction $f_{\rm flux}$ and a number recovery fraction $f_{\rm num}$ are needed. Their behavior could vary in each image, depending on the noise level as well as the degree of crowding. Flux and number recovery fractions can be obtained by core insertion experiments. A number of synthetic cores of a given flux are randomly placed into the image. Then, dendrogram is run on the new image. The fraction of the flux and number of artificial cores recovered gives the value of $f_{\rm flux}$ and $f_{\rm num}$. This is repeated for a range of fluxes, in order to obtain both $f_{\rm flux}$ and $f_{\rm num}$ as a function of flux (or, equivalently, mass).

\begin{figure*}
    \centering
    \includegraphics[width=\linewidth,trim={1cm 0 2cm 0}, clip]{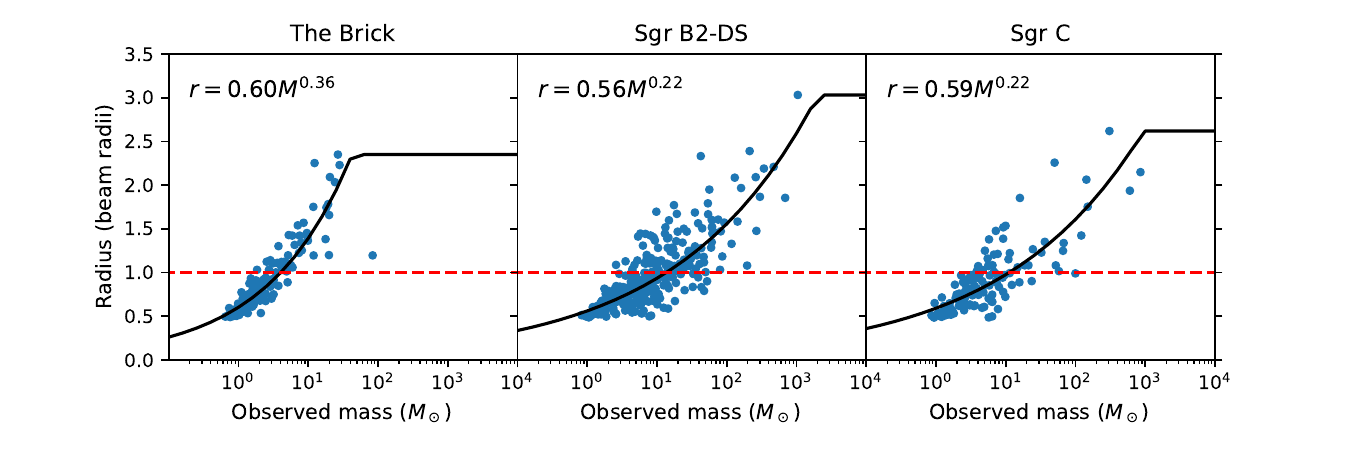}
    \caption{Core radius versus mass or raw cores detected in the three CMZ clouds. The beam radius is marked by the horizontal red dashed line. The beam FWHM is 0.94\as~for the Brick, 0.41\as~for Sgr~B2-DS and 0.69\as~for Sgr~C. Power law fits to these data are shown.}
    \label{fig:mass-radius-relation}
\end{figure*}

In previous papers (\citetalias{cheng2018}, \citetalias{liu2018}, \citetalias{oneill2021}), the synthetic cores were given the same shape as the synthesized beam, in order to represent small, unresolved cores. In this work, we insert cores with more realistic sizes. This is motivated by the observation that a significant number of the identified cores are larger than the beam. Furthermore, there is a positive correlation between estimated core mass and size (see Figure \ref{fig:mass-radius-relation}). 

\subsubsection{Radius Determination} \label{met:radius}
To determine sizes of cores, the radius calculated by the \texttt{astrodendro} package was used. \texttt{astrodendro} calculates a standard deviation of the flux distribution along the major and minor axis of the core \citep{rosolowsky2008}. The direction of the major axis is determined by principal component analysis, i.e., determining along which direction the variance in position is largest. Once the direction of the major axis has been measured, the major axis standard deviation is given by
\begin{equation}
    \sigma_{\text{maj}}^2=\frac{\sum I_i(x_{{\rm maj},i}-\Bar{x}_{{\rm maj}})^2}{\sum I_i},
\end{equation}
where $I_i$ is the intensity of pixel $i$, $x_{{\rm maj},i}$ is the pixel's position along the major axis, $\Bar{x}_{\rm maj}$ is the mean value of $x_{\rm maj}$ and the index $i$ ranges over all pixels within the dendrogram structure. The equation for $\sigma_{\text{min}}$ is analogous.
The radius is then calculated as the geometric mean of the major and minor standard deviations: $\sigma_{\text{dendro}}=\sqrt{\sigma_{\text{maj}}\sigma_{\text{min}}}$. This radius measure was chosen since it can be easily translated into the size of a synthetic Gaussian core. For an idealized Gaussian core, $\sigma_{\text{dendro}}$ equals the true standard deviation of the Gaussian.

Once the observed mass and radius of each core was determined, a power law was fitted to the points, as shown in Figure~\ref{fig:mass-radius-relation}. Note that we do not attempt beam-deconvolution of the radii. Some cores are observed to have a radius smaller than the beam, which is a consequence of the minimum core area being set to 50\% of a beam area.
The radius function was cut off at the radius of the largest observed core, to stop the inserted cores from growing to unrealistically large sizes.
The power law was then used to determine the size of the synthetic cores.

\subsubsection{Iterations}

There is a notable caveat with determining core sizes in this way, namely that the mass-radius relation is made from observed masses and radii. For an accurate core insertion experiment, we need to know the true radius of a core with a given true mass. There is no straight-forward way to correct this, since the conversion from observed to true mass requires known flux recovery fractions. To solve this issue, we took an iterative approach.

First, a power law is fitted to observed core radii as described in Section~\ref{met:radius}. Then, cores are inserted with radii given by said power law. The flux recovery and radius recovery fraction are calculated, and corrections are applied to the masses and radii of the observed cores. A new power law is fitted, but this time using the flux-corrected masses and radius-corrected radii. The process is iterated 20 times. Note that the flux and radius corrections are always applied to the observed cores, not the core properties from the last iteration. This means that the masses and radii can both increase and decrease between iterations.

An issue that was seen in early tests of the iterative method is that the flux recovery curve did not converge towards a single result, but rather oscillated between two distinct shapes. 
To mitigate this issue, a damping step was introduced into the calculation. Instead of feeding the flux recovery from one iteration directly into the next, an average was taken between the previous and new flux recovery:
\begin{equation}
f_{\rm flux,n}=\frac{f_{\rm flux,n,raw}+f_{\rm flux,n-1}}{2}.
\label{eq:damping}
\end{equation}


\subsubsection{Probability Distribution of Inserted Cores}

If synthetic cores are inserted uniformly into the image, the results may be biased depending on the amount of empty space in the image. Cores inserted on an empty background are much more likely to be detected than cores inserted in a crowded environment. To mitigate this, cores were inserted according to a probability distribution. We smoothed the ALMA image to a scale of 20\as and used it as a proxy for the probability distribution. This effectively means that cores are more likely to be inserted in regions with many other cores. The smoothing scale was chosen to be much larger than a typical core, in order to avoid inserting cores only around the few brightest sources in the image.

\begin{figure*}
    \gridline{ \fig{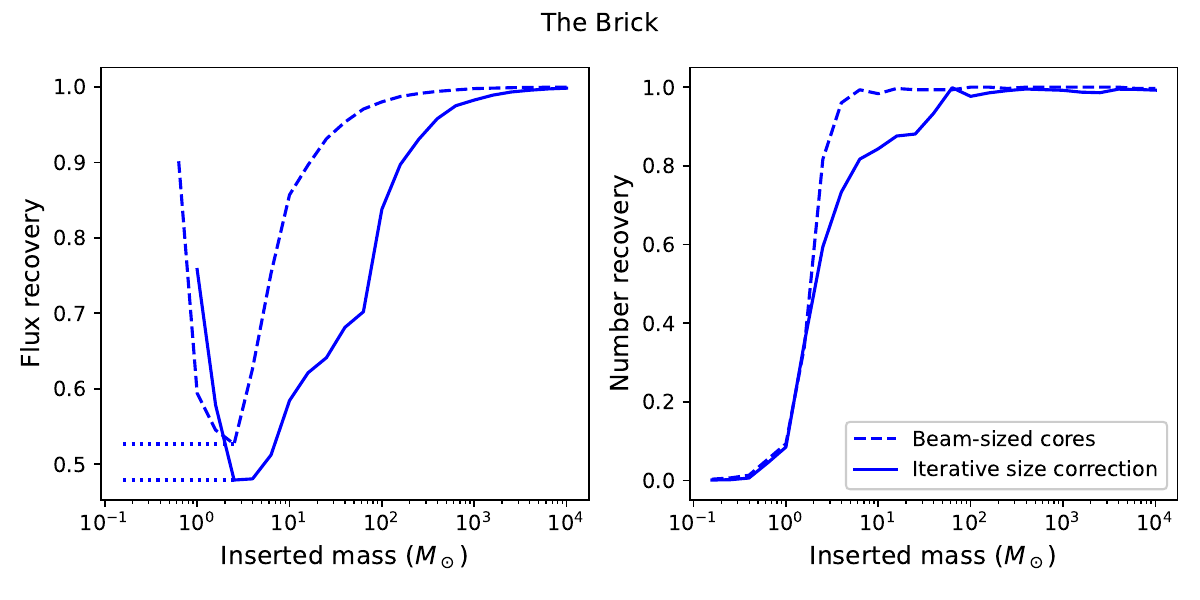}{0.75\textwidth}{}}
    \gridline{ \fig{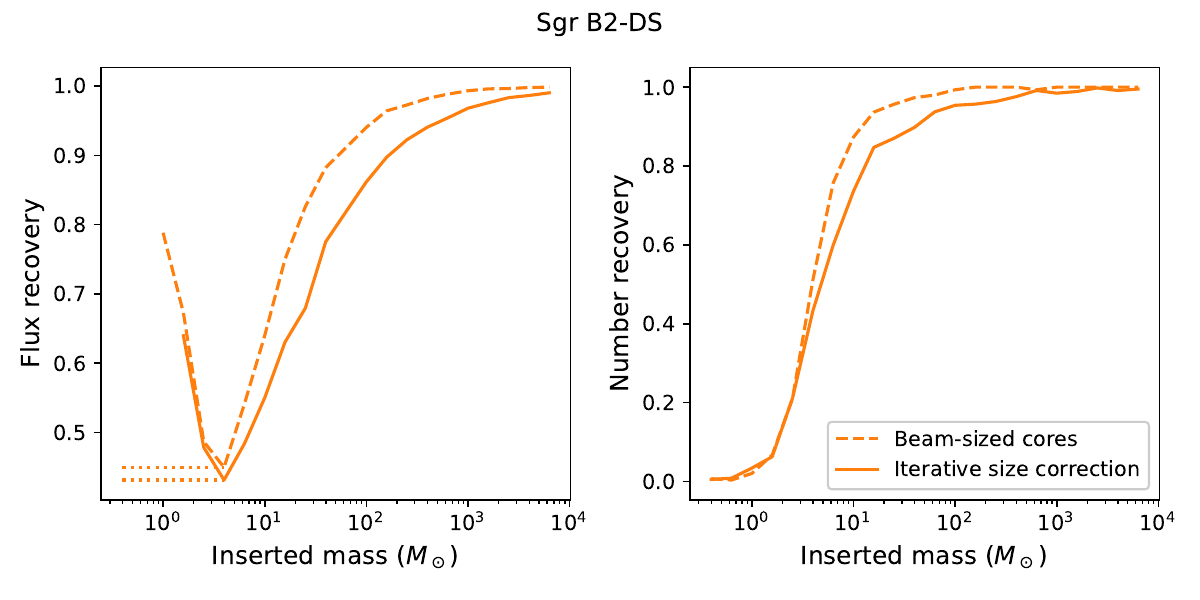}{0.75\textwidth}{}} 
    \gridline{ \fig{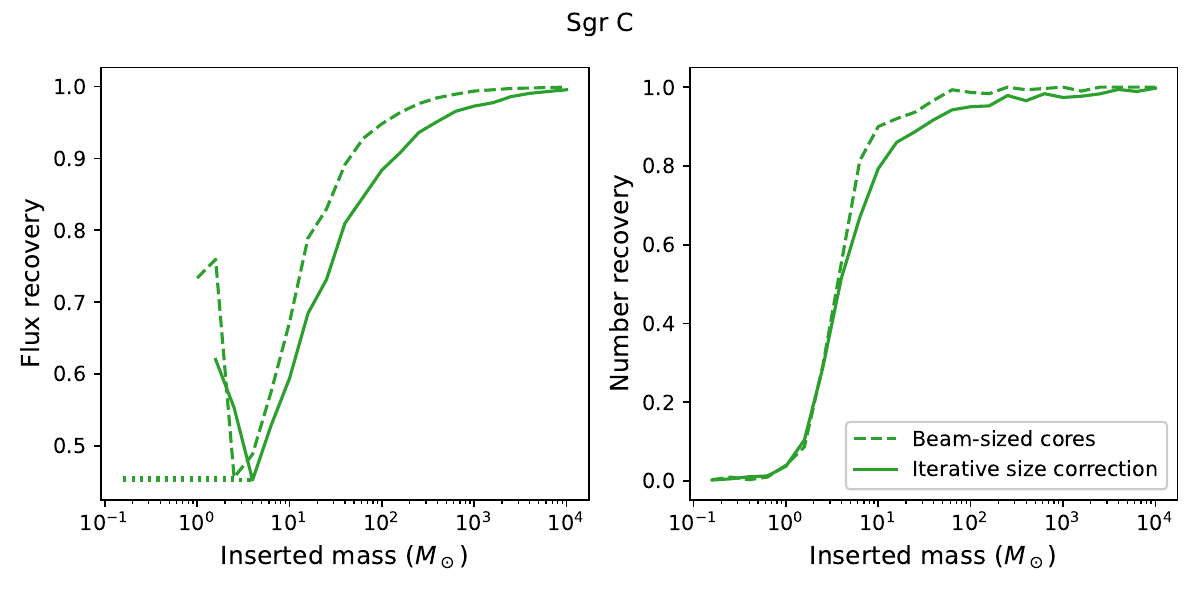}{0.75\textwidth}{}}
    \caption{Flux and number recovery fractions for the three regions in the CMZ. When allowing the size of the cores to vary with mass, both flux and number recovery decrease.}
    \label{fig:flux_number_recovery}
\end{figure*}

\subsubsection{Implementation of Core Insertion and Recovery}

Each core insertion experiment consists of inserting three circular, gaussian cores of a given flux into the original image. The cores are randomly placed according to the probability distribution described above. The number of inserted cores is kept low to avoid distorting the recovery via blended inserted cores. However, three cores are inserted at a time instead of one in order to lessen the number of experiments needed. To obtain better statistics, the experiment is repeated 100 times. The synthetic cores are not retained between experiments, each time the new synthetic cores are placed into the original image. The process is done for a range of logarithmically spaced masses, with five mass bins per decade. The bins are centered on 1 \msol, 10 \msol, 100 \msol~etc.
After each core insertion, dendrogram is run again. All new cores, i.e., those that do not have an exact correspondence among the old observed cores, are compared to the positions of inserted cores. If the peak position matches the inserted position within a tolerance of 0.28 core diameters, the core is detected. However, if the detected peak also matches with the peak of an old core, and said old core is more massive than the inserted core, the detection is discarded. This is to avoid false detections. If, e.g., a 1 \msol~core is inserted close to the peak of an existing 100 \msol~core, and the sum of the two cores is detected, it should not count as a detection of a 1 \msol~core.

The flux recovery fraction $f_{\rm flux}$ is obtained as the median ratio of recovered flux to inserted flux. Following the methods of the previous papers in this series, cores whose detected flux is larger than their true flux are not counted. These cores are considered to have falsely assigned fluxes, which could for example happen if a small core is inserted on a noise feature or the edge of a larger core. We note that including these suspected bad fluxes in the computation does not significantly change the results of the paper.

The number recovery fraction $f_{\rm num}$ is obtained as the number of recovered cores divided by the number of inserted cores.

\subsection{Recovery Functions}

The left column of Figure \ref{fig:flux_number_recovery} shows the obtained flux recovery fractions for two different core insertion methods: Insertion of beam-sized cores, similar to Papers \citetalias{cheng2018,liu2018,oneill2021}; and iterative core insertion with realistic sizes. The new method gives a lower flux recovery value than the old method. The effect is most pronounced in the Brick data. This difference is expected. With the new method, core radius increases at the same time as core mass, which means that the peak intensity of the Gaussian core increases more slowly than in the beam-sized case. In an ideal situation without noise, the flux recovery of a Gaussian core is directly determined by the peak intensity relative to the dendrogram threshold.

For both methods, large values of $f_{\rm flux}$ are obtained for the lowest mass cores. This is due to noise features being falsely identified or contributing to the cores. As in previous papers in this series, to remove this effect, masses below the mass with minimum $f_{\rm flux}$ are assumed to have constant $f_{\rm flux}$. This correction is also done within the iterative process of the realistic size core insertion.

Note that the values on the $x$-axis of Figure \ref{fig:flux_number_recovery} represent inserted, or \enquote{true} mass. In order to correct core masses, true mass must be converted to observed mass. This is done by multiplying the center mass of each bin with the corresponding flux recovery. 

The right column of Figure \ref{fig:flux_number_recovery} shows the number recovery fractions. For all three regions, number recoveries are low for masses below 1 \msol, but thereafter rise quite steeply. For the new method, the number recovery rises more slowly towards unity. Again, this is expected since the core profiles are flatter, and therefore do not stand out against the background as much as if they had been beam-sized.

To ensure satisfactory convergence of the recovery curves, convergence plots for the flux, number and radius recovery were analyzed. Systematic changes in the recovery values mainly happen during the first 10 iterations. After that, there tends to be small fluctuations in both positive and negative direction. Between 10 and 20 iterations, the recovery fraction in each mass bin fluctuates at most 0.04 from the mean. Thus, we consider the recovery curves obtained after 20 iterations to be acceptably stable, and conclude that a larger number of iterations is not needed. Example convergence plots can be found in Appendix \ref{app:convergence}.  

\begin{figure*}
    \centering
    \includegraphics[width=0.78\linewidth]{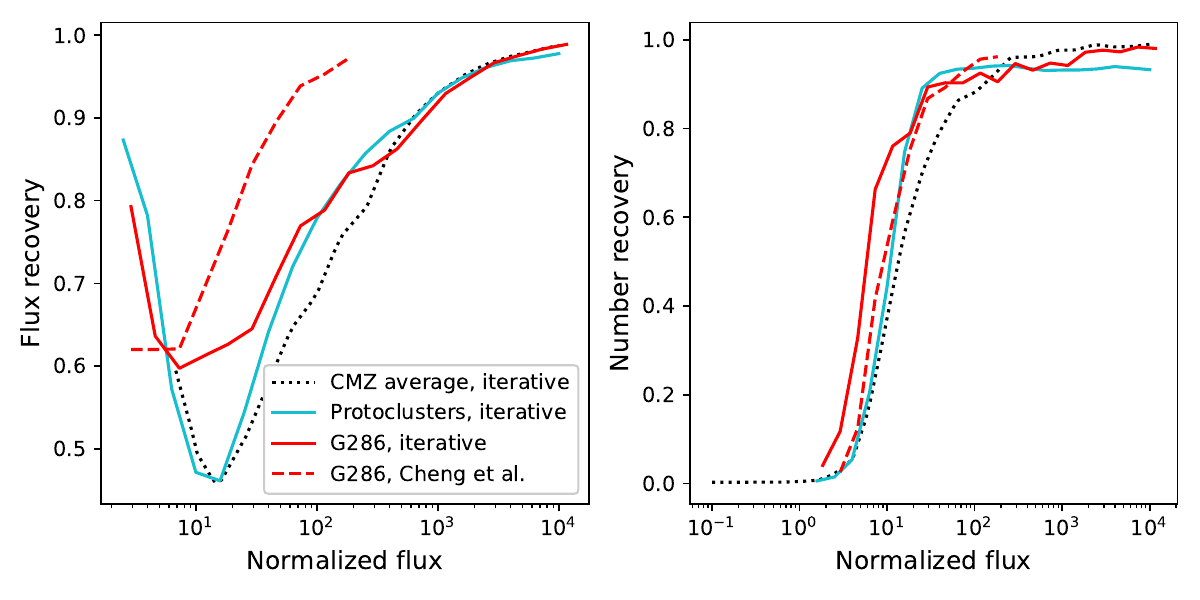}
    \caption{Flux and number corrections for G286.21+17 and the massive protocluster sample, as a function of normalized flux. The dotted line shows an average of the three CMZ regions as a comparison. The red dashed line shows the flux and number recovery obtained by \citet{cheng2018}.}
    \label{fig:old corr}
\end{figure*}

\subsubsection{Corrections to Previous CMFs}

In order to examine how our new method affects results of previous papers in the series, we apply the developed core insertion method to the regions from Paper \citetalias{cheng2018}-\citetalias{oneill2021}. The ALMA data used in Paper \citetalias{cheng2018} is a mosaic in a single region, so the method described in this paper can be applied directly. The ALMA data for the regions from Paper \citetalias{liu2018} and \citetalias{oneill2021} consists of numerous single pointings with a small number of cores detected in each, so we make a few minor changes, detailed below.

There are too few cores in each pointing to form a meaningful mass-radius relation. The pointings also have different beams, noise levels, and distances to the source, so the flux recovery fraction as a function of mass may be very different for each region.
Following Paper \citetalias{oneill2021} we estimate the recovery curves using normalized flux, defined as the flux in Jy divided by the noise level $\sigma$ expressed in Jy~beam$^{-1}$, instead of mass.

Instead of a mass-radius relation, a relation between normalized flux and radius (in terms of beam radii) was used to determine the size of the inserted cores in the ALMAGAL pointings. The flux and number recovery functions were also functions of normalized flux, rather than mass. This allowed us to combine cores from all pointings when creating the radius function and recovery curves. The recovery curves obtained from the ALMAGAL pointings were used to correct the IRDC sample from Paper \citetalias{liu2018} as well. Recovery functions for the previous regions are shown in Figure \ref{fig:old corr}.

We note that this analysis would be made more complete by smoothing the ALMAGAL images to the same linear resolution. However, since this analysis is mainly meant for qualitative evaluation of the new method, we have omitted this step.

\section{Results}\label{S:results}

\begin{figure*}
\gridline{ 
    \includegraphics[height=0.6\linewidth,trim={0.5cm 0 0 0},clip]{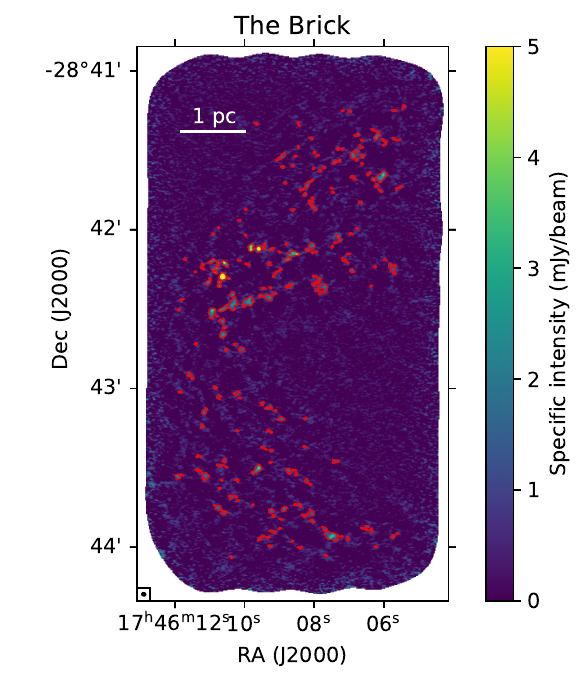}
    \includegraphics[height=0.6\linewidth,trim={0.5cm 0 0 0},clip]{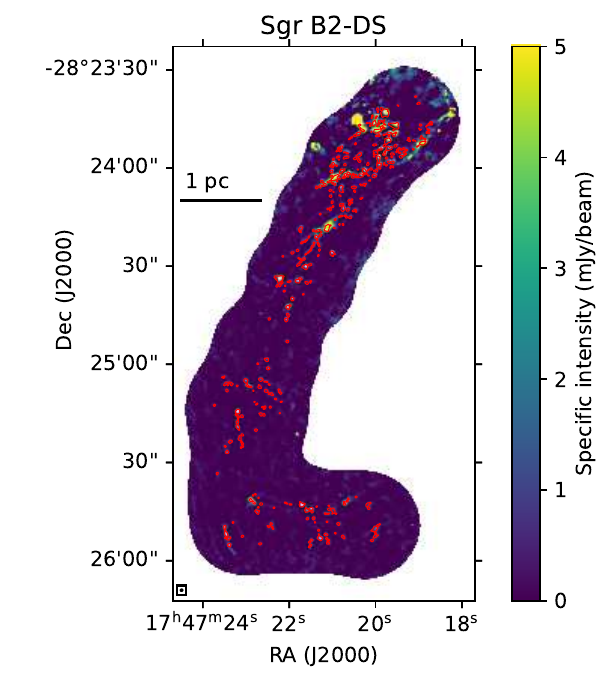} }
\gridline{
    \includegraphics[height=0.54\linewidth]{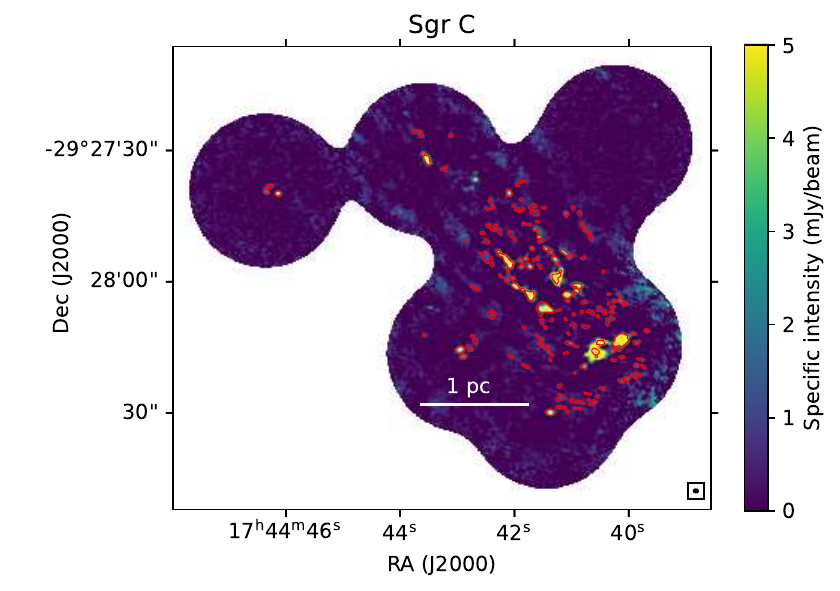} }
\caption{Detected cores in the three CMZ regions, marked with red contours. {\it (a) Top left:} The Brick. {\it (b) Top right:} Sgr~B2-DS. {\it (c) Bottom:} Sgr~C.}
\label{fig:cores_CMZ}
\end{figure*}

\subsection{Continuum Images and Core Properties}
\label{res:cores}

In Figure \ref{fig:cores_CMZ}, the mosaics of the Brick, Sgr~B2 and Sgr~C are shown, with dendrogram-identified cores marked in red. The number of cores identified were 215 for the Brick, 337 for Sgr~B2 and 159 for Sgr~C.

Due to the resolution constraints, only cores more massive than $\sim 0.8$ \msol~could be detected. The smallest mass that can theoretically be observed depends on beam size and noise level. For the dendrogram algorithm to identify a core, it must contain a minimum of one pixel at 5$\sigma$ level and $N_{\rm min}-1$ pixels at $4\sigma$ level. For the Brick, such a core would have a mass of $0.36$~\msol. After flux correction, the theoretical minimum mass increases to 0.75~\msol. Corresponding masses for Sgr~B2 and Sgr~C are 0.35~\msol~and 0.39~\msol~in raw mass, and 0.81~\msol~and 0.87~\msol~in corrected mass.

The statistics of the detected cores can be found in Table~\ref{tab:core_data}. Individual core properties are listed in 
Table~\ref{tab:detailed_core_data}. The complete table is available in machine-readable form. Notably, the cores in Sgr~B2 and Sgr~C are generally more massive than in the Brick.
This holds true even when the images are convolved to the same (largest) beam size (see Section ~\ref{sec:resolution}).

\begin{deluxetable}{lccc}
    \tablewidth{0.45\textwidth}
    \tablecaption{Summary of Core Properties \label{tab:core_data}} 
\tablehead{ \colhead{Quantity} & \colhead{The Brick} & \colhead{Sgr~C} & \colhead{Sgr~B2-DS}}          
\startdata   
        $N_{\rm cores}$ & 215 & 159 & 377 \\
        $M_{\rm min}$, raw & 0.45  & 0.51 & 0.44 \\
        $M_{\rm min}$, corr. & 0.95  & 1.13 & 1.02 \\
        $M_{\rm max}$, raw & 83.79  & 601.96 & 756.73 \\
        $M_{\rm max}$, corr. & 100.01  & 623.62 & 787.67\\
        $M_{\rm median}$, raw & 1.46  & 3.07 & 6.29\\
        $M_{\rm median}$, corr. & 3.04  & 5.92 & 11.07\\ 
        $\Sigma_{\rm mm,median}$ & 0.20 & 0.60 & 2.56\\
\enddata
\tablecomments{Masses are listed in \msol, and are presented both before flux correction (raw) and after (corr.). Mass surface density, $\Sigma$, is given in g cm$^{-2}$.}
\end{deluxetable}

\begin{deluxetable*}{cccccccccc}[h]
        \tablewidth{\textwidth}
    \tablecaption{Sample Core Data Table  \label{tab:detailed_core_data}} 
\tablehead{
  \colhead{ID} &  \colhead{$\ell$} & \colhead{$b$} & \colhead{$I_{\rm peak}$} &  \colhead{$F_{\nu}$} &  \colhead{$M_{\rm c,raw}$}  &  \colhead{$M_{c}$}  &    \colhead{$R_{c}$} & \colhead{$\sigma_{\text{dendro}}$} &  \colhead{$\Sigma_{c}$}  \\ [-1ex]
 \colhead{}   & \colhead{($\degr$)}  & \colhead{($\degr$)} & \colhead{(mJy $\rm beam^{-1}$)} &  \colhead{(mJy)} &  \colhead{$(M_{\odot})$} &  \colhead{$(M_{\odot})$}  & \colhead{(0.01 pc)} & \colhead{(0.01 pc)} &   \colhead{(g $\rm cm^{-2}$)}   
 }
 \startdata
 Brick.c1  & 0.260970  & 0.016159 & 40.20             & 60.09     & 83.79       & 100.01    & 5.29     & 1.92          & 1.99                            \\
 Brick.c2  & 0.231768 & 0.011749 & 3.79             & 20.16     & 28.11       & 41.16     & 7.86     & 3.58          & 0.30                             \\
 Brick.c3  & 0.261278 & 0.035903   & 3.79             & 17.35     & 24.18       & 35.96     & 6.98     & 3.27          & 0.33                            \\
 Brick.c4  & 0.257326  & 0.017097 & 2.79             & 16.26     & 22.67       & 33.97     & 6.93     & 3.38          & 0.31                            \\
 Brick.c5  & 0.257932  & 0.015596 & 2.64             & 14.66     & 20.43       & 30.98     & 6.96     & 3.36          & 0.28                            \\
 Brick.c6  & 0.259142  & 0.023763  & 4.29             & 14.30      & 19.94       & 30.32     & 5.27     & 2.66          & 0.48                            \\
 Brick.c7  & 0.261493 & 0.020899 & 9.00              & 14.21     & 19.81       & 30.15     & 4.63     & 1.93          & 0.61                            \\
 Brick.c8  & 0.258349 & 0.013246 & 3.73             & 13.84     & 19.30        & 29.46     & 6.16     & 2.86          & 0.34                            \\
 Brick.c9  & 0.241813  & 0.008779 & 4.04             & 13.06     & 18.21       & 27.99     & 5.87     & 2.80           & 0.35                            \\
 Brick.c10 & 0.262007  & 0.020272 & 5.06             & 12.68     & 17.68       & 27.27     & 4.83     & 2.22          & 0.50 
 \enddata
 \tablecomments{Cores are ordered by descending mass. Coordinates are given for the centroid of the core, calculated by astrodendro. $R_c$ is the radius of a circle with the same total area as the core, while $\sigma_{\text{dendro}}$ is the astrodendro radius defined in Section \ref{met:radius}. The complete table is available in machine-readable form.}
\end{deluxetable*}

\subsection{The Core Mass Function (CMF)}

\subsubsection{Construction of the CMF}\label{res:cmfcon}

\begin{figure*}
    \centering
    \includegraphics[width=\linewidth,trim={1cm 0 2cm 0},clip]{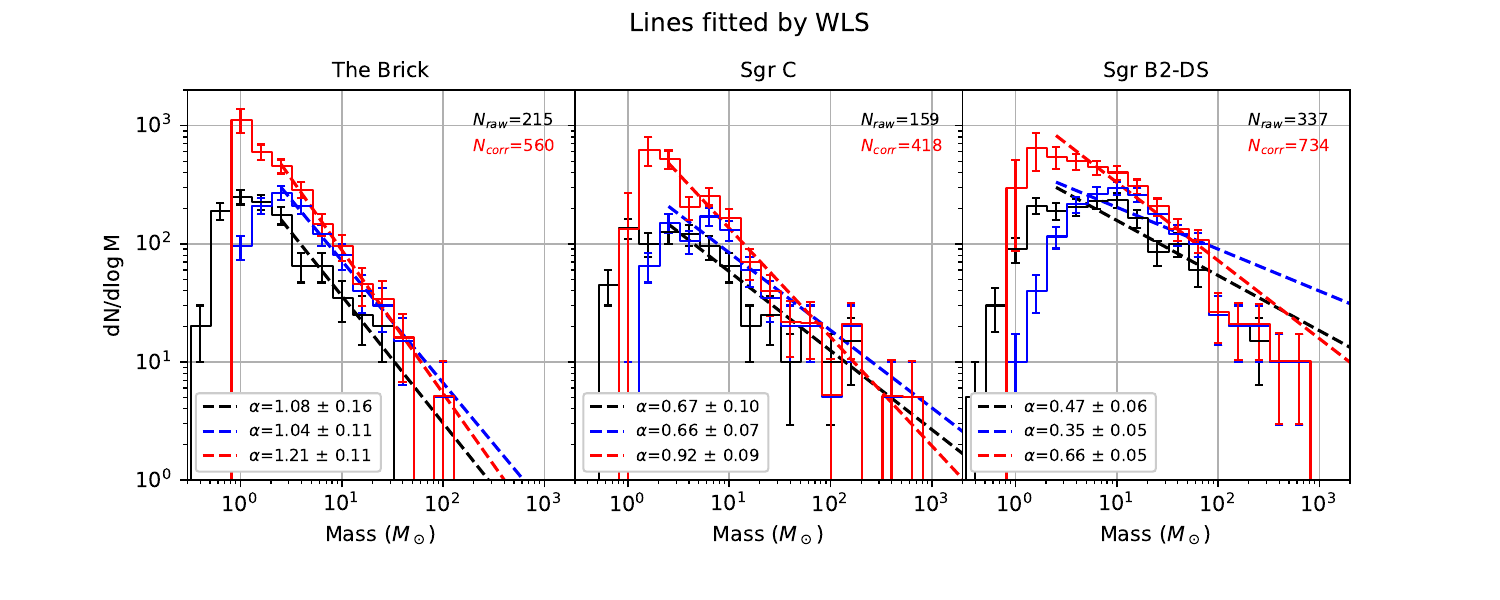}
    \caption{CMFs for The Brick (left), Sgr~C (middle) and Sgr~B2 (right). The black histogram shows the \enquote{raw} CMF, the blue histogram shows the flux-corrected CMF and the red histogram shows the number-corrected ``true'' CMF. Lines are fitted using the weighted least squares method.}
    \label{fig:CMF_CMZ_WLS}
\end{figure*}

We calculate \enquote{raw}, flux-corrected, and number-corrected (``true'') CMFs for all three regions in the CMZ, which can be seen in Figure \ref{fig:CMF_CMZ_WLS}. The binning is the same as in Paper \citetalias{cheng2018}, \citetalias{liu2018} and \citetalias{oneill2021}: the bins are evenly spaced logarithmically with 5 bins per decade, and setting one bin to be centered on 1 \msol. The error bars on the raw and flux-corrected CMF denote $\sqrt{N}$ Poisson counting errors, while the errors on the number-corrected CMF are set to be the same relative size as on the flux-corrected CMF. Note that these do not take the uncertainty in $f_{\rm flux}$ or $f_{\rm num}$ into account.

As can be seen, number correction has a dramatic effect on the low mass end of the core mass function, but a negligible effect on the high mass end. Flux correction on the other hand has an impact on intermediate to high mass bins as well. This is in contrast to the previous method of completeness correction, as is discussed further in Section~\ref{res:method_comparison}.

\subsubsection{Single Power-Law Fits to the CMF}\label{res:fitting}

\begin{figure*}
    \centering
    \includegraphics[width=\linewidth,trim={1cm 0 2cm 0},clip]{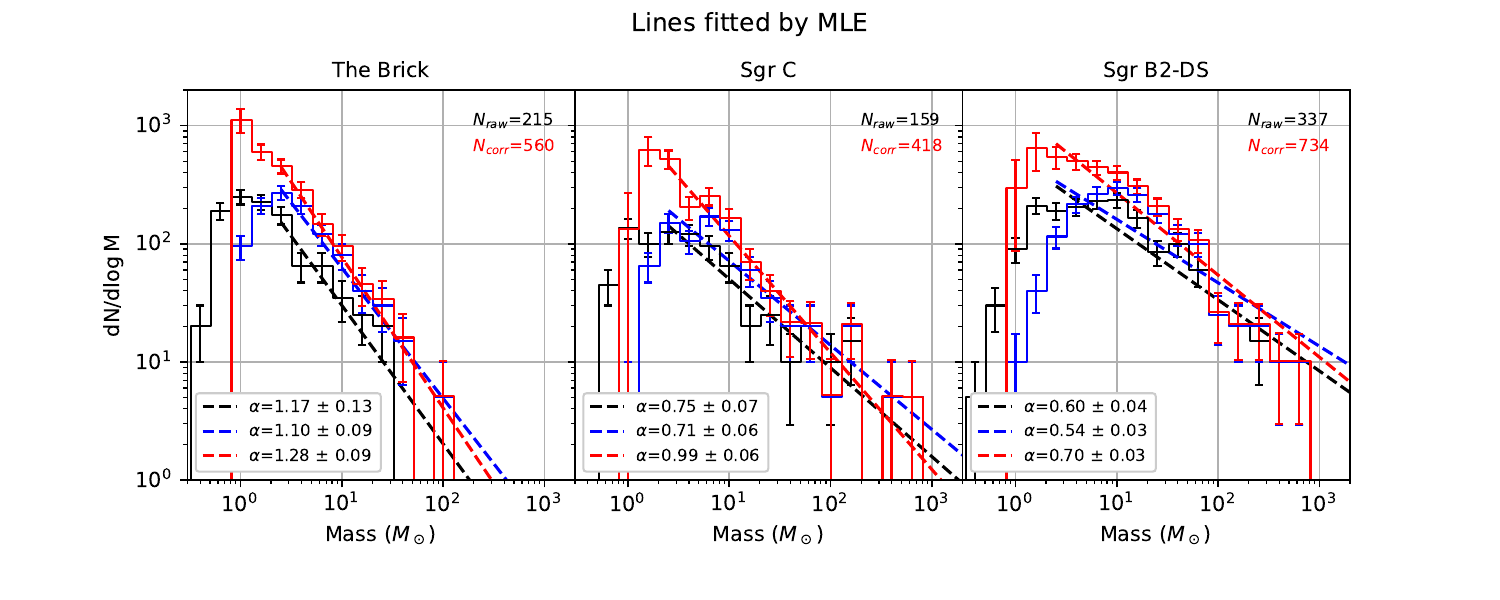}
    \caption{CMFs for The Brick (left), Sgr~C (middle) and Sgr~B2 (right). The black histrogram shows the raw CMF, the blue histogram shows the flux-corrected CMF and the red histogram shows the number-corrected CMF. Lines are fitted using the maximum likelihood estimator method.}
    \label{fig:CMF_CMZ_MLE}
\end{figure*}

A power law of the form given in Equation \ref{eq:powerlaw} is fitted to each CMF. First, this is done using the weighted least squares (WLS) method from Paper \citetalias{cheng2018}, \citetalias{liu2018} and \citetalias{oneill2021}. A least squares fit is performed in logarithmic space, with errors given by the average of the asymmetric Poisson errors. Empty bins are treated in the same way as in previous papers: the data point is assumed to be a factor 10 lower than if the bin contained one core, and the error bar reaches up to the one core level. The results are insensitive to reasonable variations in the treatment of the empty bins.
In Papers \citetalias{cheng2018}, \citetalias{liu2018} and \citetalias{oneill2021}, the power law was fitted over a standard range starting at the bin centered at 1 \msol. However, due to the higher mass sensitivity threshold of the CMFs in this paper, we instead fit from the bin centered at 2.5~\msol. 

Power law indices of the WLS fits for the CMZ regions can be seen in Figure \ref{fig:CMF_CMZ_WLS} and Table \ref{tab:plawparams}.
While the true CMF of the Brick has a power law index of $\alpha=1.21 \pm 0.11$, which is consistent with the Salpeter value of 1.35, Sgr~B2 and Sgr~C have considerably shallower indices. Their true CMFs have power law indices of $\alpha=0.66\pm0.05$ and $\alpha=0.92 \pm 0.09$ respectively.

Following Paper \citetalias{oneill2021}, we also fit power laws using a maximum likelihood estimator (MLE), which has been shown to be more accurate than WLS fits \citep{clark_generalizations_1999,white_estimating_2008}. 
A method for fitting power laws to unbinned samples was proposed by \citet{clauset2009}, using the maximum likelihood estimator \citep{newman_power_2005}
\begin{equation}
   \hat{\alpha} = n\left[ \sum_{i=1}^{n} \textrm{ln} \frac{M_{i}}{M_{\textrm{min}}} \right]^{-1},
\end{equation}
where $M_{\rm{min}}$ is the start of the fitting range, $M_{i}$ are the masses of the cores that have masses $\geq M_{\rm{min}}$  and $n$ is the number of cores with $M>M_{\rm{min}}$. Since we are fitting from the bin centered at 2.5~\msol, $M_{\rm{min}}$ is set to 2~\msol. The standard error on $\alpha$ is $\Delta\hat{\alpha} = \hat{\alpha} / \sqrt{n}$.
The MLE above requires individual core masses, not binned data. That means that it is not applicable to the number-corrected CMF, which is only defined by the bin height. Therefore, a second MLE is needed. Following Paper \citetalias{oneill2021}, we use the MLE for binned data proposed by \citet{virkarclauset2014} (hereafter called MLE-B). In the case when the bins can be written on the form $(c^s, c^{s+1}, ... c^{s+k})$ (i.e., logarithmically spaced bins), the MLE for the index is
\begin{equation}
    \hat{\alpha}=\log_c \left[ 1+\frac{1}{s-1-\log_c b_{\rm min}+(1/n)\sum_{i={\rm min}}^k ih_i} \right],
\end{equation}
with the standard error
\begin{equation}
    \Delta \hat{\alpha}= \frac{  c(c^{\hat{\alpha}}-1)  }{ c^{(2+\hat{\alpha})/2} \ln{c}\sqrt{n}   }.
\end{equation}
Here $b_{\rm min}$ represents the minimum bin and $h_i$ represents the number of counts in each bin in the sample.

The results of the MLE and MLE-B fits can be seen in Figure \ref{fig:CMF_CMZ_MLE} and Table \ref{tab:plawparams}. As in Paper \citetalias{oneill2021}, the MLE fit results in slightly steeper power laws, but the difference in true CMF is within one standard error from the WLS values. We obtain a Salpeter-like slope $\alpha = 1.28 \pm 0.09$ for the Brick, and smaller values $\alpha=0.99\pm0.06$ and $\alpha=0.70\pm 0.03$ for Sgr~C and Sgr~B2-DS respectively. Since the MLE method is known to be superior to WLS, we consider these power law indices as our main results.

We have examined the impact of varying the lower limit of the fitted mass range on the derived power law indices of the ``True'' CMFs. While the indices of the Brick and Sgr~C are relatively insensitive to the choice of fitting range, this is not the case for Sgr~B2-DS. 
For example, if the MLE-B fit is instead performed starting from the 10~\msol~bin, a steeper power law index of $1.00\pm 0.06$ is obtained. However, we again emphasize that it is desirable to have a common, simple metric that can be compared among all the regions, which we have chosen to be the single power law fit down to $\sim 2\:M_\odot$.

\begin{deluxetable}{ccccc}
\centering
\tablecaption{Best-Fit Power Laws \label{tab:plawparams}}
\tablehead{\colhead{Region} & \colhead{CMF}& \colhead{$\alpha$, WLS} &	\colhead{$\alpha$, MLE}&		\colhead{$\alpha$, MLE-B} }
\startdata  
Brick & Raw & $1.08\pm 0.16$ & $1.17\pm0.13$&	$1.18\pm 0.07$	\\
... & Flux-corr.& $1.04\pm 0.11$&	$1.10\pm0.09$ &	$1.13\pm0.09$	\\
... & True& $1.21\pm0.11$&	... &	$1.28\pm0.09$	\\
\hline						
Sgr~C & Raw& $0.67\pm0.10$ &	$0.75\pm0.07$ &	$0.75\pm0.07$	\\
... & Flux-corr. & $0.66\pm0.07$ & $0.71\pm0.06$ & $0.72\pm 0.06 $	 \\
... & True & $0.92\pm0.09$ & ... & $0.99\pm0.06$ \\
\hline						
Sgr~B2-DS & Raw & $0.47\pm0.06$ &	$0.60\pm0.04$ &	$0.60\pm0.07$	\\
... & Flux-corr.& $0.35\pm 0.05$ &	$0.54\pm0.03$ &	$0.54\pm0.03$	\\
... & True & $0.66\pm 0.05$ &	...&	$0.70\pm0.03$	
\enddata
\tablecomments{\enquote{True} refers to the number-corrected CMF.}
\end{deluxetable}

\subsection{CMFs at Common Resolution}
\label{sec:resolution}

\begin{figure*}
    \centering
    \includegraphics[width=\linewidth]{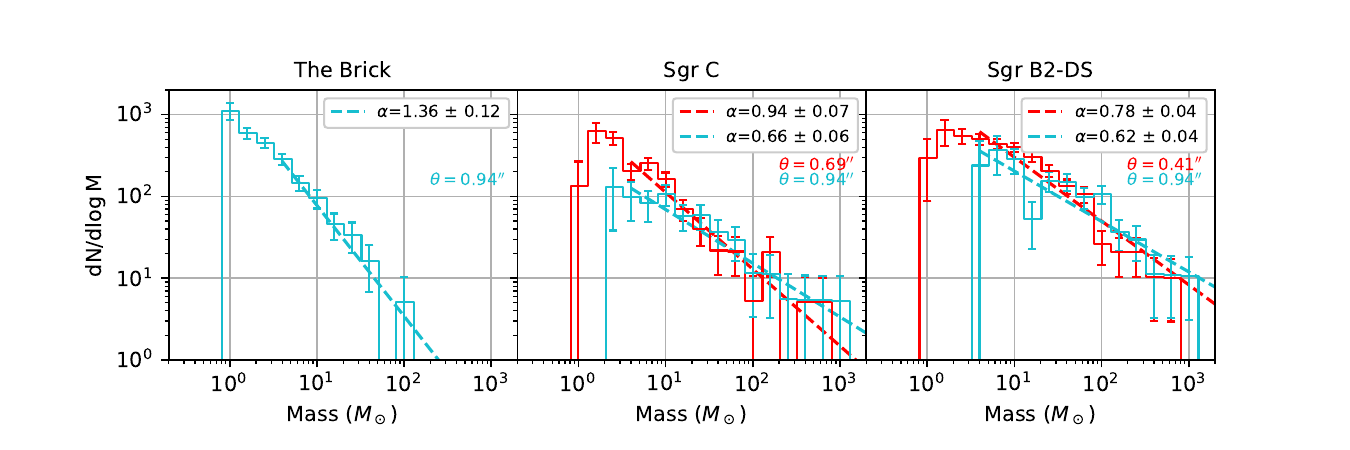}
    \caption{``True'' (number corrected) CMFs with original resolution (red) and the same resolution as the Brick (cyan). Since the low-resolution Sgr~B2-DS CMF lacks cores in the bin centered at 2.5 \msol, the fits are instead performed from the bin above (centered at 3.9 \msol). Lines are fitted by the MLE-B method.}
    \label{fig:resolution}
\end{figure*}

The spatial resolution of our images differs by approximately a factor of two between the Brick and Sgr~B2-DS. As seen in Paper \citetalias{cheng2018}, this can affect the derived index of the CMF. To investigate this effect, we smoothed the Sgr~C and Sgr~B2-DS images to a circular beam with the same area as the beam of the Brick image. Then, all CMF analysis was repeated.
Figure \ref{fig:resolution} shows the resulting number-corrected CMFs. Note that our standard fitting bin, centered at 2.5~\msol, is empty in the Sgr~B2-DS smoothed CMF. Therefore, we start all fits from the bin above instead.  The smoothed versions of the CMFs have shallower indices than the original CMFs: $\alpha = 0.66\pm 0.06 $ for Sgr~C and $\alpha = 0.62\pm 0.04$ for Sgr~B2-DS. These can be compared with the Brick power law index $\alpha = 1.36 \pm 0.12$, which is obtained when fitting from the same bin. In other words, the index of the Brick CMF is significantly steeper than the other two regions. This result could also be seen at the original resolution.
On the other hand, the power law indices of Sgr~C and Sgr~B2-DS become quite similar when smoothed to uniform resolution. We conclude that the steeper CMF index in the Brick cannot be explained by the lower resolution of its data, but that the relative values of Sgr~C and Sgr~B2-DS may be affected.

\subsection{Comparison Between Correction Methods}
\label{res:method_comparison}

\begin{figure*}
    \centering
    \includegraphics[width=0.8\linewidth,trim={0.5cm 1cm 1cm 1cm}, clip]{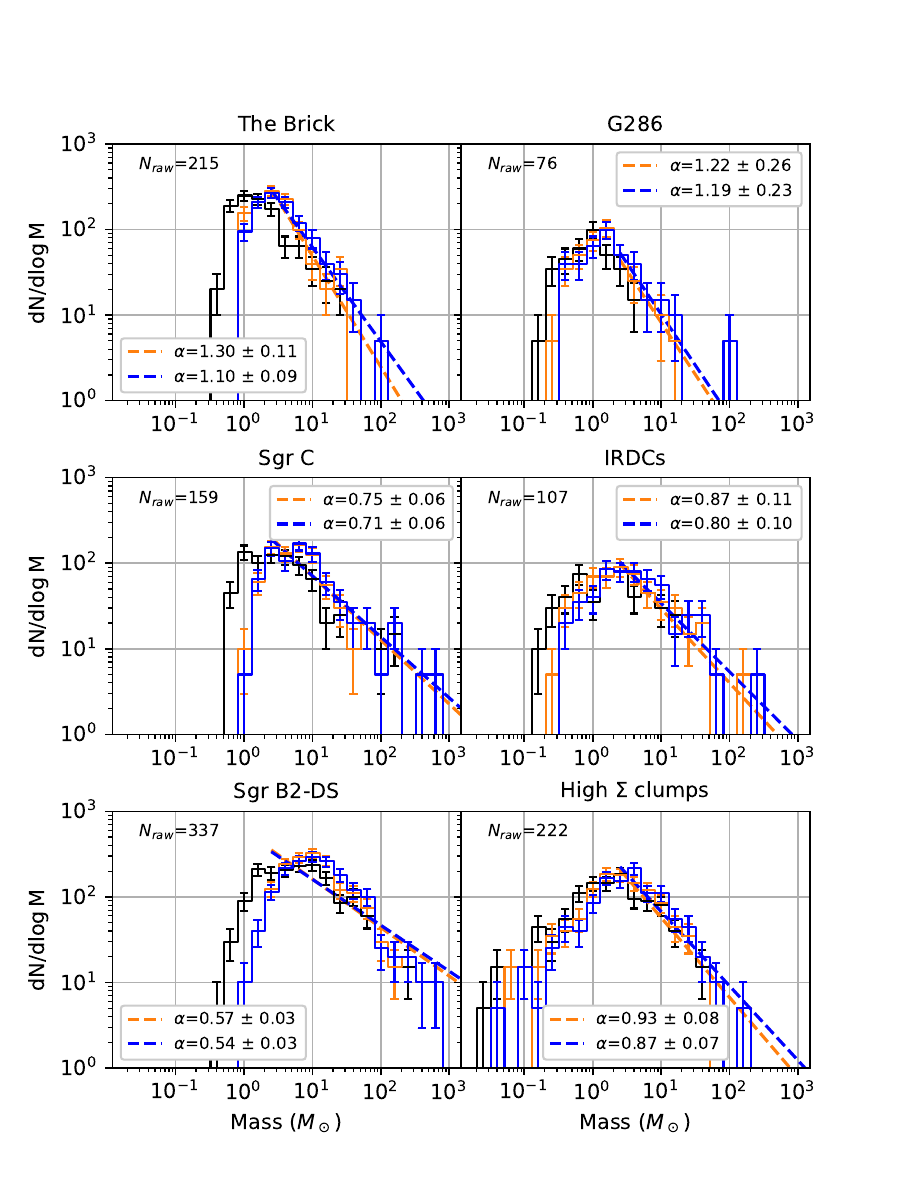}
    \caption{Comparison of raw and flux-corrected CMFs for the two core insertion-recovery methods. The black histogram shows the raw CMF. The orange histogram shows the flux-corrected CMF using core insertion with beam-sized cores. The blue histogram shows the flux-corrected CMF using realistically-sized cores, presented in this work.
    Power laws are fitted via the MLE method starting from 2 \msol. Left column: Regions in the CMZ. Right column: Regions studied in Paper \citetalias{cheng2018}, \citetalias{liu2018} and \citetalias{oneill2021}: from top to bottom: G286.21+0.17, IRDCs, and massive protoclusters from the ALMAGAL survey.} 
    \label{fig:fluxcorr_CMF}
\end{figure*}

\begin{figure*}
    \centering
    \includegraphics[width=0.8\linewidth,trim={0.5cm 1cm 1cm 1cm}, clip]{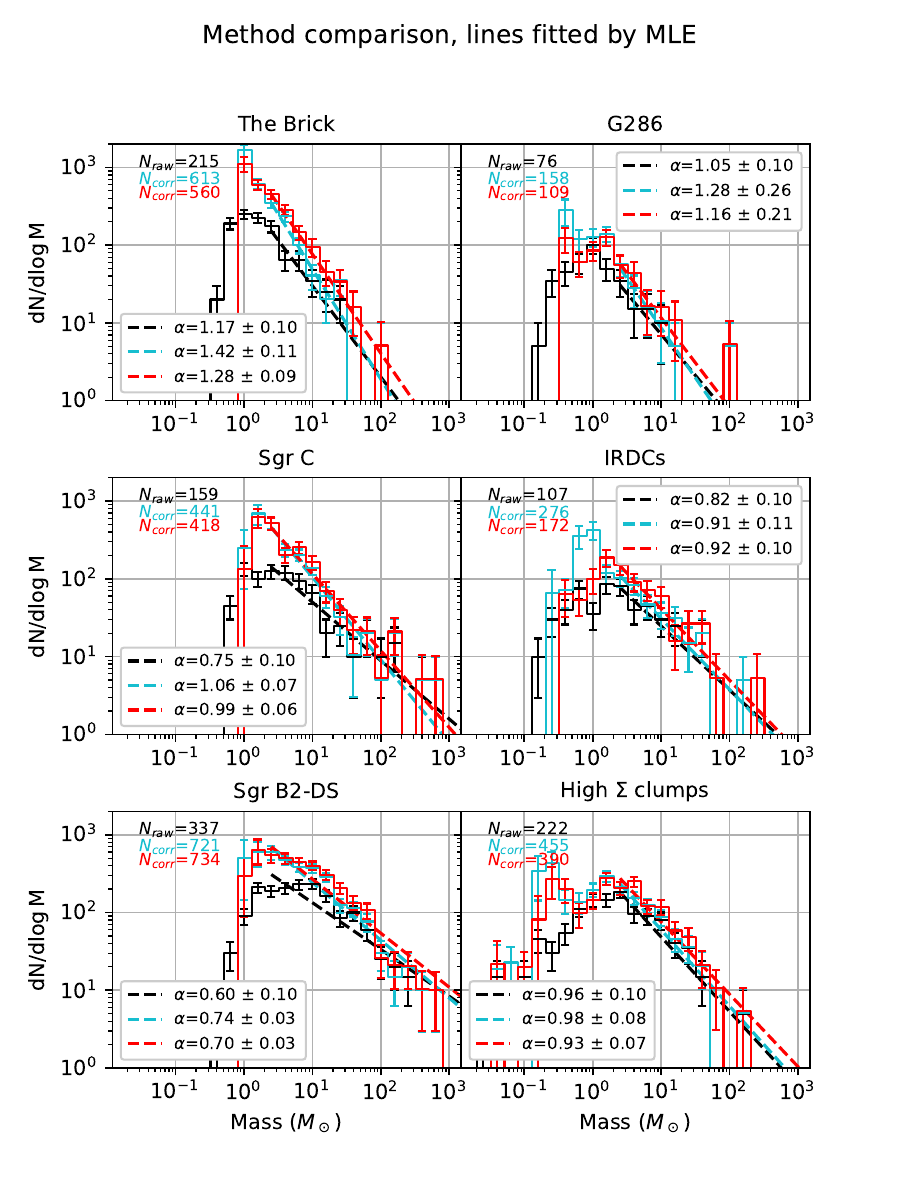}
    \caption{Comparison of raw and number-corrected CMFs for the two core insertion-recovery methods. The black histogram shows the raw CMF. The cyan histogram shows the number-corrected CMF using core insertion with beam-sized cores. The red histogram is the number-corrected CMF using realistically-sized cores, presented in this work. Power laws are fitted via the MLE method (raw CMF) and MLE-B method (true CMFs) starting from 2 \msol. Left column: Regions in the CMZ. Right column: Regions studied in Paper \citetalias{cheng2018}, \citetalias{liu2018} and \citetalias{oneill2021}: from top to bottom: G286.21+0.17, IRDCs, and massive protoclusters from the ALMAGAL survey.}
    \label{fig:numcorr_CMF}
\end{figure*}

In this section, we investigate the effect on the CMF caused by using the completeness correction method described in this paper using realistically sized cores for insertion-recovery, as opposed to the method used in Paper \citetalias{cheng2018}-\citetalias{oneill2021} using beam-sized cores. To ensure comparability between CMF indices, we re-calculate the power law fits to the CMFs from previous papers, using the MLE/MLE-B method and starting from the bin centered at 2.5 \msol.

In Figure \ref{fig:fluxcorr_CMF}, the flux-corrected CMF is shown for both the old and new methods. In Figure \ref{fig:numcorr_CMF}, the number-corrected CMF is shown for the two methods. For the CMZ regions, the \enquote{old} corrections are derived by inserting beam sized cores into the image (see dashed curves in Figure \ref{fig:flux_number_recovery}). For the other regions, the old CMFs are taken directly from the respective papers.\footnote{However, in the case of the massive protocluster CMF from Paper \citetalias{oneill2021}, masses have been recalculated due to discovery of a previous error in the mass assignments of cores.} 
We note the new method appears to shift the flux-corrected CMF to higher mass than the old method. This is especially visible for the Brick and G286.21+0.17, shown in the top left and top right panels. This is because the new method generally gives lower flux recovery fractions over the whole mass range, and therefore the corrections have a larger effect. In all the core samples, the new flux-corrected CMF has a shallower index than the old one.

Differences in the number-corrected CMF are shown in Figure \ref{fig:numcorr_CMF}. Note that the new version of the number-corrected CMF tends to have slightly higher values towards the high mass bins than the old method. The effect is most clearly visible in the Brick (top left), G286.21+0.17 (top right) and the IRDC sample (middle right). The indices of the number-corrected CMFs are, however, quite robust under the change of method. As with the flux-corrected CMFs, there is a trend towards shallower CMFs with the new method. However, the change in $\alpha$ is less than or comparable to 1$\sigma$ standard error in all cases.

The peak of the CMF may be more influenced by the new method than the power law index. As can be seen in the CMFs from Paper \citetalias{liu2018} and \citetalias{oneill2021} (middle and bottom right in Figure \ref{fig:numcorr_CMF}), the large peaks present in the old CMFs become significantly smaller with the new method. This is likely because the flux-correction factor for the smallest cores increases with the new method, moving the least massive cores upwards in the mass scale. The low-end of the number correction factor is however very modestly affected by the new method, meaning that the smallest cores are moved into a mass range with a lower degree of number correction than before.

\subsection{Statistical Comparison of CMFs}
An Anderson-Darling (A-D) test was performed to determine if any of the CMFs analyzed in this paper are similar enough to be sampled from the same underlying distribution. 
The test is especially sensitive to differences in the tail of the distributions. The $k$-sample A-D test uses  the test statistic
\begin{equation}
    A_{kN}^2 = \frac{1}{N} \sum_{i=1}^k \frac{1}{n_i}\sum_{j=1}^{N-1}\frac{(NM_{ij}-jn_i)^2}{j(N-j)},
\end{equation}
where $n_i$ is the number of elements in the $i$:th sample. If we define the ordered list $Z$ containing all elements from all samples, $M_{ij}$ is the number of elements in sample $i$ that are smaller or equal to the j:th element of $Z$. $N$ is the total number of elements \citep[see, e.g.,][]{scholz1987}. In our case, $k=2$ since we are comparing the CMFs pairwise.
We have used the implementation from the python package \textit{scipy}\footnote{\url{https://docs.scipy.org/doc/scipy/reference/generated/scipy.stats.anderson_ksamp.html}}.
A lower value of $A_{kN}^2$ means larger similarity between the distributions. From the value of $A_{kN}^2$ and the number of elements in each sample, a $p$-value can be calculated. Since the number-corrected CMF is defined by bin only, we perform bootstrap resamplings according to the method in 
Paper \citetalias{liu2018} to derive a mean $p$-value. In each mass bin, $N_{\rm bin}$ uniformly distributed masses are generated. $N_{\rm bin}$ is the number of cores in the bin according to the number-corrected CMF. The generated CMFs of two regions are compared with the A-D test. The procedure is repeated 3000 times, and the mean $p$-value is calculated.

In Table \ref{tab:AD}, the $p$-values obtained from the A-D test can be seen for each pair of CMFs. A low $p$-value indicates that the samples are likely drawn from different distributions. We choose $p<0.005$ as the limit of statistical significance. The CMFs were compared at native resolution. To avoid any influence from different mass sensitivities, only the cores exceeding 2 \msol~(consistent with the lowest fitting bin) were included.

The ``true'', number corrected CMF of Sgr~B2-DS is found to be different from all other regions. We also see that the Brick is different from the massive protocluster sample, but we cannot rule out similarity between the Brick and any of the other regions.
The CMFs of the previously analyzed regions cannot be distinguished from each other or from Sgr~C. However, note that the previous regions have relatively small numbers of cores above 2 \msol, which decreases the power of the A-D test. If a lower limit of 0.79 \msol~is applied instead, most of the CMF pairs are significantly different, for example, the G286 and massive protocluster samples as well as the Brick and Sgr~C. 

We also performed an A-D test at common resolution for the regions in the Galactic center. This test included cores above 3.2 \msol, since this is the lowest populated bin in the smoothed Sgr~B2-DS true CMF. In that case, the Brick was significantly different from both Sgr~C and Sgr~B2, with $p<0.001$. For Sgr~C and Sgr~B2 on the other hand we obtained $p>0.25$, and they could thus not be distinguished by the test.

\begin{figure*}
    \centering
    \includegraphics[width=\linewidth]{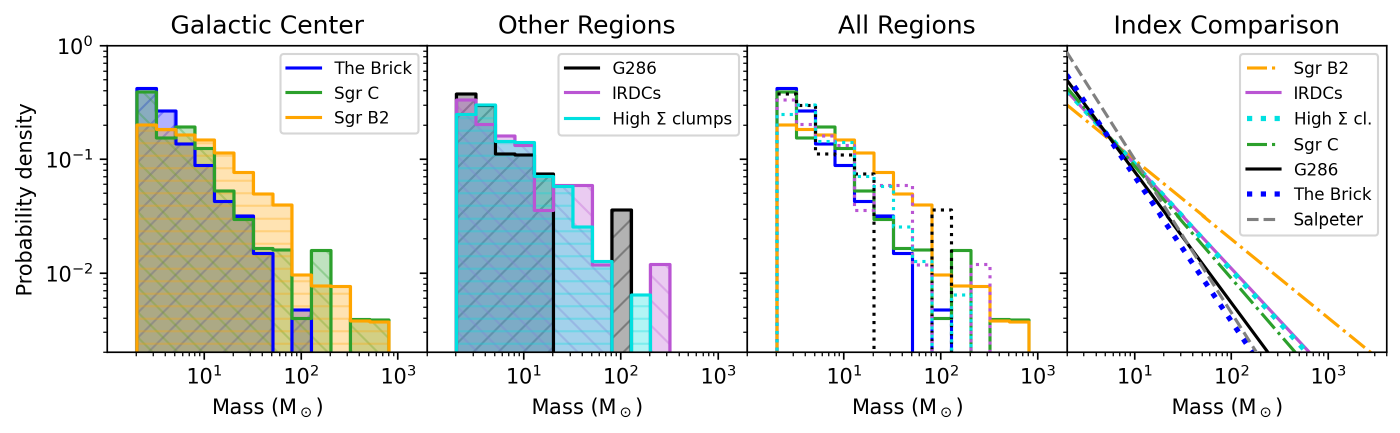}
    \caption{The first three panels show true core mass functions, excluding cores with masses below 2 \msol~and normalized by the total number of cores. Panel 1: CMZ regions. Panel 2: regions from Paper \citetalias{cheng2018}-\citetalias{oneill2021}. Panel 3: All CMFs together (regions from previous papers are shown with dotted lines). Panel 4 illustrates the differences between the CMF power law fits (best fit from 2 \msol~using the MLE-B method). }
    \label{fig:PDF_comparison}
\end{figure*}

The normalized, native-resolution core mass functions used for the A-D test are shown in Figure \ref{fig:PDF_comparison}. In the first panel, the differences between the three CMZ distributions can be seen. The second panel shows CMFs for the regions studied in Paper \citetalias{cheng2018}-\citetalias{oneill2021}. The most notable difference is that G286 
has a smaller fraction of cores more massive than 20~\msol\ compared to the other regions. In the third panel, all CMFs are plotted together. Finally, the fourth panel shows the indices of all true CMFs, as well as the Salpeter value.

 \begin{deluxetable*}{ccccccc}
 \tablewidth{0.45\textwidth}
 \tablecaption{$p$-values of A-D Test on Number-Corrected CMFs \label{tab:AD}} 
\tablehead{ \colhead{...} & 
\colhead{Sgr~B2-DS} & \colhead{Sgr~C} & \colhead{G286} & \colhead{IRDCs} & \colhead{High $\Sigma$ clumps}
}             
\startdata  
        The Brick & \textcolor{red}{$<0.001$} & $0.03$ &  $>0.05$ & $0.02$ & \textcolor{red}{$0.001$}   \\
        Sgr~B2-DS & ... & \textcolor{red}{$<0.001$} & \textcolor{red}{$0.002$} & \textcolor{red}{$0.002$} & \textcolor{red}{$0.001$}  \\
        Sgr~C & ... & ... & $>0.05$ & $>0.05$ & 0.05 \\ 
        G286 & ... & ... & ... & $>0.05$ & $>0.05$ \\ 
        IRDCs & ... & ... & ... & ... & $>0.05$ \\ 
\enddata     
\tablecomments{Only cores above 2 \msol~are included. $p$-values below 0.005 are marked in red, indicating that the distributions are significantly different.}
\end{deluxetable*}

%

\subsection{Dense Gas Fraction} \label{res:dense_gas}

We estimate the dense gas fraction of the three CMZ regions, which we define as the total mass of the cores divided by the total mass of the region. Note that different definitions can be found in the literature, so comparisons between works should be done with caution. To calculate the total core mass, we use the number-corrected CMF. The mass of the cores in each bin is estimated as the number of cores multiplied by the central mass of the bin. 

Furthermore, we estimate the total mass of each region using 1.1~mm continuum images from the Bolocam Galactic Plane Survey, version 2 \citep{ginsburg2013_BOLOCAM}. The mass surface density is calculated over the entire ALMA footprint, using Equation \eqref{eq:Sigmamm}. The mass surface density is then converted to mass using Equation \eqref{eq:coremass}. Here $T_d=20$ K is assumed and $\kappa_\nu$ is obtained in the same way as for the core mass calculation. Note that cores are not detected in the entire ALMA field of view, but only in the region where the primary beam response is above 0.5. Ideally, we would want to match the area over which we measure the mass to the area where we detect cores. However, we want to avoid using a patch from the Bolocam data that is significantly smaller than the beam. Since the Sgr~B2 map in particular covers a thin strip, restricting to primary beam response $>0.5$ gives a region that is thinner than the Bolocam beam FWHM by a factor of $\sim 2$. A similar issue arises with the Sgr~C map. Therefore, the estimated dense gas fractions are likely underestimations.

In Table~\ref{tab:dense_gas}, total core masses, large-scale masses estimated from Bolocam data and dense gas fractions are shown for the three regions. The Brick has the lowest dense gas fraction, with only 3~\% of its mass contained in dense cores. Sgr~B2-DS has a dense gas fraction of 14~\%, while Sgr~C has a value of 24~\%. We note that there may be systematic errors in our dense gas fraction values. The core mass and total mass are estimated by different instruments and at slightly different wavelengths.   
When we compare our Bolocam-derived masses to Herschel-masses from \cite{battersby2020}, we find that their masses for the Brick and Sgr~C-Dense are larger by a factor $\sim$1.5.
However, systematic uncertainties should not strongly affect our regions relative to each other. Differences in maximum recoverable scale between the ALMA images could affect how much flux is recovered, but we note that the region with the largest maximum recoverable scale (the Brick), which should recover most flux at the core scale, is also the region with the lowest dense gas fraction. If all regions had the same maximum recoverable scale, the difference between the Brick and the others would only increase.

\begin{deluxetable}{cccc}
\tablewidth{0.45\textwidth}
\tablecaption{Core Masses, Total Masses and Dense Gas Fractions\label{tab:dense_gas}} 
\tablehead{ \colhead{Region} & 
\colhead{Core Mass} & \colhead{Total Mass} & \colhead{Dense Gas} \\[-2ex]
 & 
\colhead{($10^{3}$\msol)} & \colhead{($10^{3}$\msol)} & \colhead{Fraction} }
\startdata
The Brick & $1.8$ & $54$& 0.03 \\
Sgr C & $4.0$ & $17$ & 0.24\\
Sgr B2-DS & $11$ & $80$ & 0.14
\enddata
\end{deluxetable}

\subsection{Spatial Distribution of Cores}\label{res:spatial}

\begin{figure*}
\gridline{
    \includegraphics[height=0.45\linewidth]{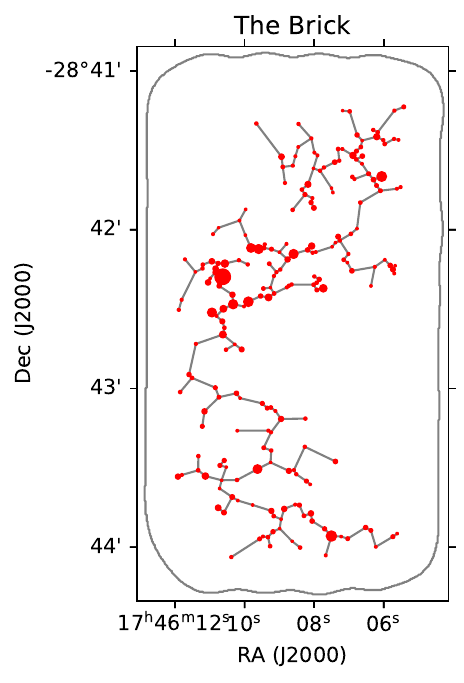}
    \includegraphics[height=0.45\linewidth]{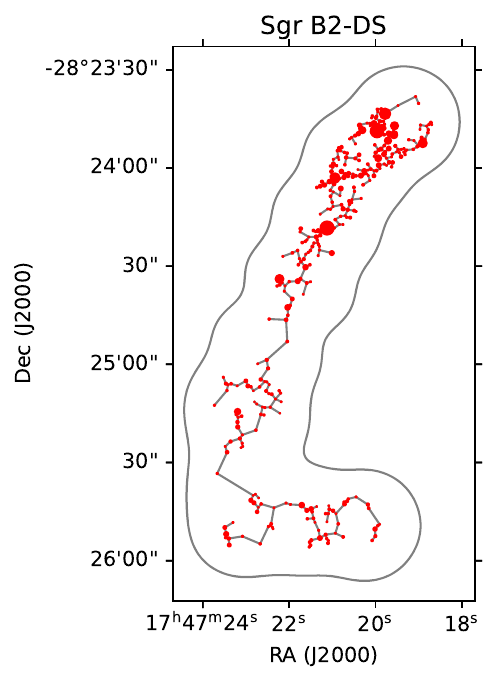}
    }
\gridline{
    \includegraphics[height=0.4\linewidth]{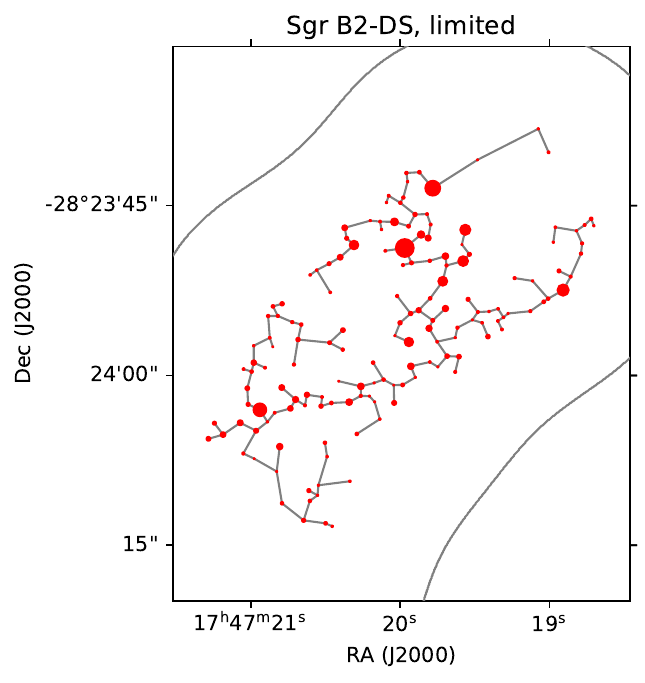}
    \includegraphics[height=0.4\linewidth]{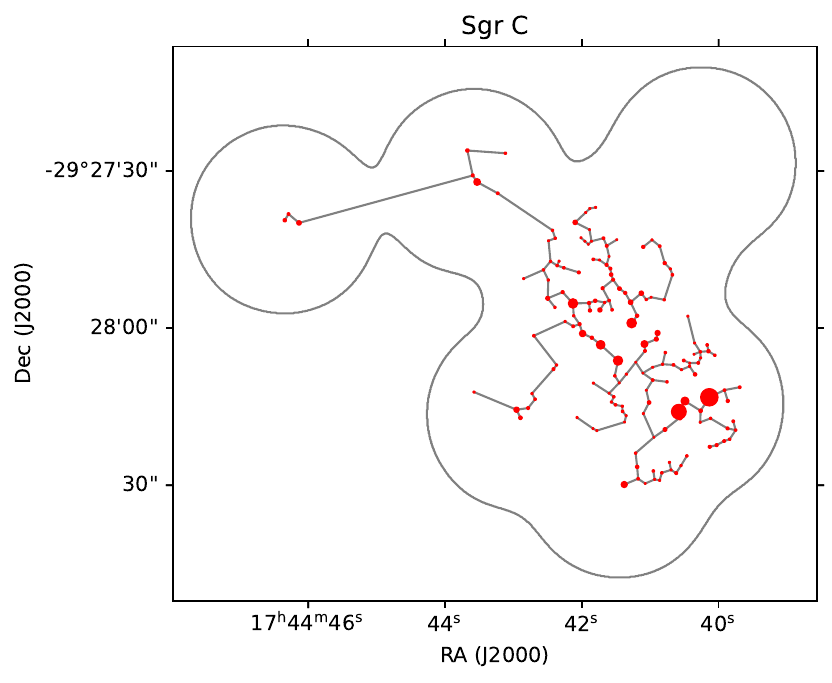} 
    }
\caption{Minimum spanning trees for the CMZ regions. The circular symbols represent the cores identified by this work, and their area is proportional to the estimated core mass. Note that the area normalization differs between regions. }
\label{fig:MST}
\end{figure*}

In order to quantify the spatial distribution of cores in the CMZ regions, the $Q$ parameter was calculated. The $Q$ parameter was developed by \citet{cartwright2004} and is defined as
\begin{equation}
Q = \frac{\bar{m}}{\bar{s}},
\end{equation}
where $\bar{m}$ is the normalized mean length of the minimum spanning tree (MST) of the cores, and $\bar{s}$ is the normalized mean separation between cores (i.e., the mean of all distances between cores). A spanning tree is a set of edges connecting all cores without any cycles, and the minimum spanning tree is the spanning tree that minimizes total edge length. Note that the normalizations of $\bar{m}$ and $\bar{s}$  are different: 
\begin{equation}
    \bar{m}=\frac{m (N_{\rm cores}-1)}{\sqrt{N_{\rm cores}\pi R_{\rm cluster}^2}},
\end{equation}
where $N_{\rm cores}$ is the number of cores and $R_{\rm cluster}$ is the cluster radius. On the other hand,
\begin{equation*}
    \bar{s}=\frac{s}{R_{\rm cluster}}.
\end{equation*}

The resulting minimum spanning trees can be seen in Figure~\ref{fig:MST}. The size of the symbols is proportional to the mass of the core. The $Q$ parameters for each region are listed in Table \ref{tab:Q-parameter}. All $Q$ parameters are below 0.8, which indicates that the regions are substructured rather than radially concentrated. The most strongly substructured region is Sgr~B2 with $Q=0.32$, followed by the Brick with $Q=0.52$ and Sgr~C with $Q=0.71$. However, the $Q$ parameter was developed for clusters that are approximately circular in projection, and can be biased if the region deviates too strongly from that. The map of Sgr~B2 is very elongated, which can explain its low $Q$ value. To remove this effect, the $Q$ parameter is also calculated for the north end of the Sgr~B2 map, where the aspect ratio has been limited to 2:1. Note that this region (seen in Figure \ref{fig:MST}, lower left) contains 153 cores, which is close to half of the cores in the Sgr~B2 map. The obtained $Q$ value for the limited region is significantly higher, $Q=0.67$. Thus we conclude that the precise value of the Sgr~B2 $Q$ parameter is sensitive to the definition of the global region.

\begin{deluxetable}{ccccc}
\tablewidth{0.45\textwidth}
\tablecaption{Core Clustering Properties\label{tab:Q-parameter}} 
\tablehead{ \colhead{Region} & \colhead{$\bar{m}$} & 
\colhead{$\bar{s}$} & \colhead{$Q$}
}             
\startdata  
        The Brick & 0.37 & 0.71 & 0.52 \\
        Sgr C & 0.27 & 0.37  & 0.71\\
        Sgr B2-DS & 0.21 & 0.63 &0.32\\
        Sgr B2-DS, lim. & 0.43 & 0.65 & 0.67 \\
         \hline
\enddata    
\end{deluxetable}

\subsection{Mass Segregation of Cores}\label{res:massseg}

\begin{figure*}
    \centering
    \includegraphics[width=\linewidth,trim={1cm 0 2cm 0},clip]{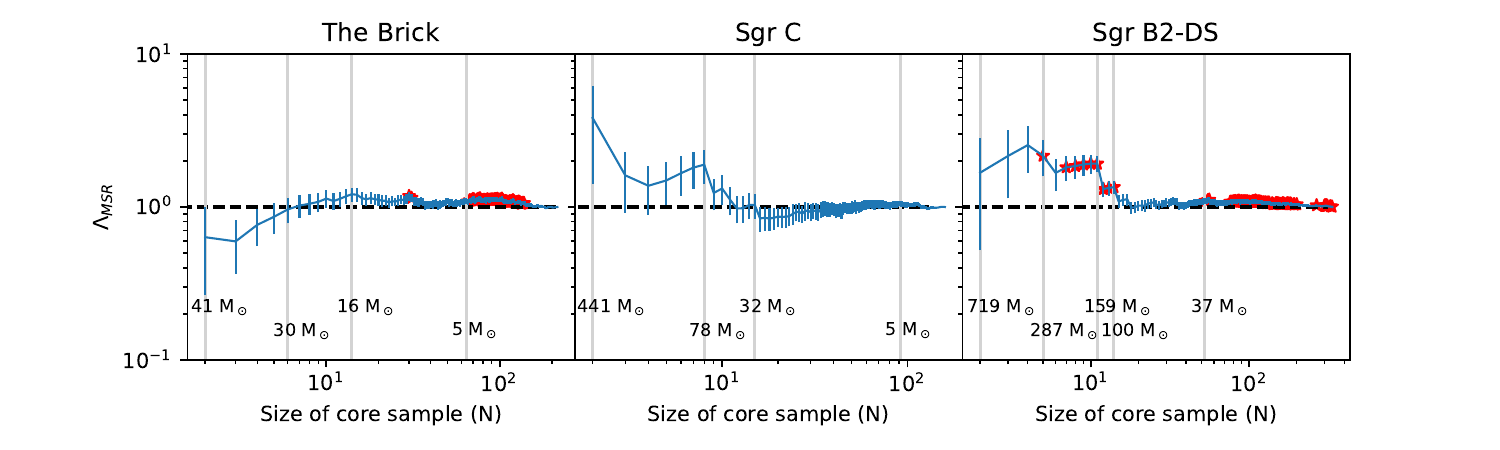}
    \caption{Mass segregation parameter $\Lambda_{\rm MSR}$ as a function of core number $N$. Red stars mark values that deviate from 1 by more than two standard deviations.}
    \label{fig:lambda_msr}
\end{figure*}

We quantified the mass segregation in the regions using the mass segregation parameter $\Lambda_{\rm MSR}$ \citep{allison2009}. Just like the $Q$ parameter, $\Lambda_{\rm MSR}$ is calculated using minimum spanning trees. It is defined as
\begin{equation}
    \Lambda_{\rm MSR}=\frac{<l_{\rm random}>}{l_{\rm massive}},
\end{equation}
where $<l_{\rm random}>$ is the average total length of the MST of $N$ randomly chosen cores, and $l_{\rm massive}$ is the total MST length of the $N$ most massive cores. If the cluster is mass segregated, we expect the most massive cores to be closer together than a group of randomly selected cores, giving $\Lambda_{\rm MSR}>1$. An inversely mass segregated cluster, where the most massive cores are more spaced out than other cores, would have $\Lambda_{\rm MSR}<1$. More extreme segregation yields values further from unity.

Mass segregation parameters for the CMZ regions can be seen in Figure \ref{fig:lambda_msr}. Here $<l_{\rm random}>$ was calculated using 1000 iterations for $N\leq20$, and 100 iterations for larger $N$ values. For the Brick, $\Lambda_{\rm MSR}$ is close to 1 for all sample sizes. Even though $\Lambda_{\rm MSR}$ is significantly different from 1 below the scale of 5 \msol, the size of the difference is small enough to be negligible. Sgr~B2 on the other hand, shows significant mass segregation for a range of masses between 100 and 290 \msol. There appears to be two levels of segregation. In the approximate range 160-290 solar masses we have $\Lambda_{\rm MSR} \sim 1.7-2.2$, while the mass range 100-160 \msol~shows a lower segregation ($\Lambda_{\rm MSR} \sim 1.3$). There are also a number of statistically significant values below 37 \msol. However, these $\Lambda_{\rm MSR}$ values are all $\lesssim 1.1$, which is so close to 1 that we do not consider it evidence for physically meaningful mass segregation. Finally, the graph for Sgr~C shows elevated values of $\Lambda_{\rm MSR}$ going up to $\sim 1.9$ for the eight most massive cores. However, the difference from 1 is just below the threshold of statistical significance. The evidence for mass segregation is thus weaker than in Sgr~B2, even though the magnitude of the proposed segregation is similar.

\begin{figure}
    \centering
    \includegraphics[width=0.9\linewidth]{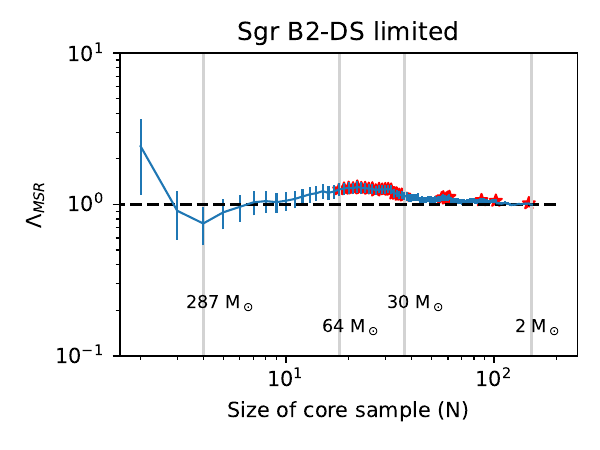}
    \caption{Mass segregation parameter $\Lambda_{\rm MSR}$ as a function of core number $N$, for a limited part of the Sgr~B2-DS map. Red stars mark values that deviate from 1 by more than two standard deviations. The maximal significant mass segregation is $\Lambda_{\rm MSR}\sim 1.3$.}
    \label{fig:lambda_msr_lim}
\end{figure}

It should be noted that most of the massive cores in Sgr~B2-DS are located in the northern part of the map. This fact is responsible for most of the mass segregation. We also calculate mass segregation within the limited region used for the $Q$ parameter calculation (see Figure \ref{fig:MST}, bottom left). The result can be seen in Figure \ref{fig:lambda_msr_lim}. These cores still show significant mass segregation, but only in the range 30-60 \msol~and with a lower maximum value of 1.3.

\section{Discussion}
\label{S:discussion}

\subsection{Previously Estimated Core Numbers}

Here we compare our core populations to those reported by other papers studying these regions.

\citet{ginsburg2018} detected 271 cores in Sgr~B2 using 3 mm continuum images with a resolution of approximately 0.5\arcsec. Of these, 89 are within the region analyzed in this paper. We note that this is significantly fewer than we find, but such differences are to be expected due to the differing methods. \citet{ginsburg2018} identified sources by eye rather than by an automated algorithm. By comparing core positions, we find that all 89 cores except one have a corresponding core in this work (although there are a few examples of two \citet{ginsburg2018} cores corresponding to a single core in this work). \citet{ginsburg2018} do not probe down to the same mass scale as our work: i.e., their least massive core has an estimated mass of 7.8 \msol (or about 16~\msol\ if a dust temperature of 20~K is used, as is done in our paper). Furthermore, \citet{jeff2024} used the same ALMA data as this work to specifically search for hot cores in Sgr~B2-DS. Nine such cores were found, as further discussed in Section \ref{dis:caveats}.

\citet{lu2020} used the same ALMA dataset as this paper to derive a core mass function in Sgr~C. They identified 275 cores, which is almost double the number in our work.  Their reported median mass (1.8 \msol) is also lower than the median mass derived by us (3.1 \msol). However, the ALMA project contains two sets of observations with different antenna configurations. \citet{lu2020} have combined the observations to obtain a resolution of 0.25\as $\times$ 0.17\as . However, we have only used the more compact configuration data, since its resolution is more similar to the other regions we have analyzed. This difference in resolution is most likely the reason for the difference in reported core numbers. We note that the \citet{lu2020} methods of core identification are similar to ours, with modestly different dendrogram parameters. However, they did not perform completeness corrections. Even though a direct comparison of CMF power law indices may not be meaningful due to these differences in methods, \citet{lu2020} also derived a slightly top-heavy shape of the CMF in Sgr~C, reporting a power-law index  $\alpha=1.00\pm0.13$ starting from $\sim 6$~\msol.

\begin{figure}
\gridline{\fig{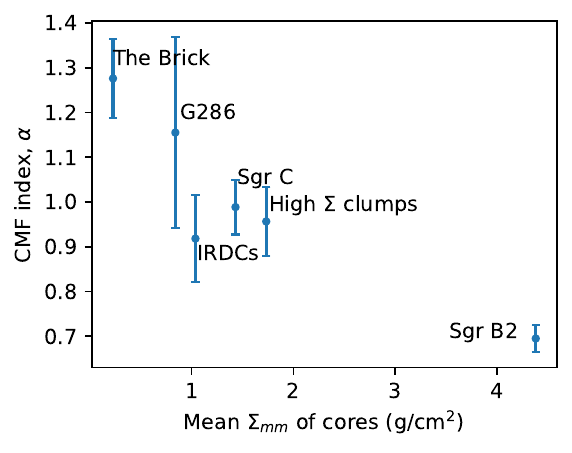}{0.95\linewidth}{}}
\gridline{\fig{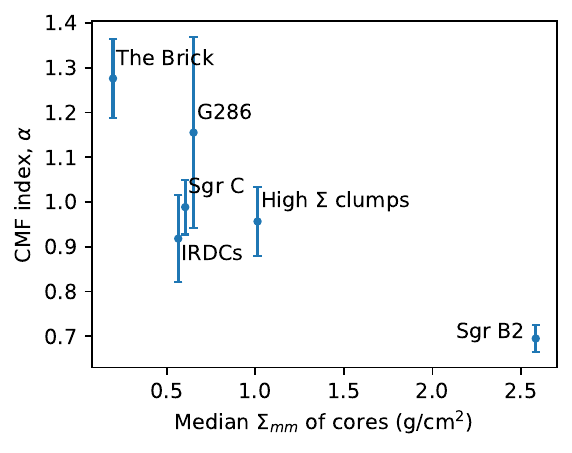}{0.95\linewidth}{}}
\caption{Relation between CMF power law index (MLE-B fit from 2 \msol) and average mass surface density of the cores. Top: CMF index as a function of mean core $\Sigma_{\rm mm}$. Bottom: CMF index as a function of median core $\Sigma_{\rm mm}$.}
\label{fig:msd_vs_slope}
\end{figure}

\subsection{Relation Between CMF Power Law Index and Evolutionary Stage}
\label{dis:evolutionary_stage}

We have calculated the core mass function in three regions in the CMZ, in order to investigate how the CMF varies with environment. 
The CMFs of the three CMZ regions were found to be significantly different from each other, with the Brick having the steepest index and Sgr~B2 the shallowest. The same pattern is seen both at common and original resolution. This could be related to the evolutionary stage of the regions. The dense gas fraction in the Brick (0.03) is significantly lower than in Sgr~B2-DS (0.14) and Sgr~C (0.24), which indicates an earlier evolutionary stage. Since Sgr~B2-DS and Sgr~C have dense gas fractions of the same order, we cannot say with certainty that one is more evolved than the other. When comparing our dense gas fractions to the values reported in \cite{battersby2020} (defined as the fraction of mass in 0.1-2 pc scale structures), we find a similar relation between the Brick and Sgr~C, which strengthens our conclusion that the Brick is in an earlier evolutionary stage than Sgr~C.

Furthermore, the evolutionary stage can be indicated by the star formation rate. The Brick shows few signs of star formation \citep{immer2012,mills2015}, and its star formation rate is estimated to $10^{-4}-10^{-3}$ \msol~yr$^{-1}$ \citep{lu2019,henshaw2023}.
Both Sgr~B2-DS and Sgr~C have been found to harbor massive star formation \citep[e.g.,][]{ginsburg2018,lu2019}. The star formation rate in the entire Sgr~B2 cloud has been estimated in the range $\sim 0.04-0.08$\msol~yr$^{-1}$ \citep{kauffmann2017,ginsburg2018,lu2019}, while the star formation rate in Sgr~C has been found to be $\sim0.8 - 2.7 \times 10^{-3}$ \msol~yr$^{-1}$  \citep{kauffmann2017,lu2019}. Although the global SFR in Sgr~B2 is much higher than in Sgr~C, Sgr~B2 is also much more massive. The specific SFR, i.e. star formation rate divided by cloud mass, is therefore similar for the two clouds. Using a mass of $1.4\times10^{6}$ \msol~for Sgr~B2 and $2.4\times10^{4}$ \msol~for Sgr~C \citep{kauffmann2017}, we obtain specific SFRs that are within a factor of 2 of each other. Previous studies have shown shallow core mass functions in high density regions and regions with massive star formation \citep[e.g.,][]{motte2018,kong2019,pouteau2022,pouteau2023}. Our results are in agreement with these previous findings.

There are different ways to interpret such a correlation. One possibility is that the regions that favor massive star formation (for example due to high density) may also favor the development of a top-heavy CMF. Another possibility is that the index of the CMF is directly related to the evolutionary stage of the cores. This view is supported by \cite{nony2023}, who found that prestellar cores in W43 had a steeper CMF than protostellar cores in the same region. The protostellar cores were also more massive in general. This indicates that the CMF index becomes more shallow as the cores of the region evolve. Cores may thus be accreting more material during their lifetime \citep[see, e.g.,][]{sanhueza2019}. However, results from the previous papers in this series compel us to add some nuance to this view. The G286 region is known to be relatively evolved, but still has a CMF index close to the Salpeter value. The IRDCs from Paper \citetalias{liu2018} on the other hand are in an earlier evolutionary stage, yet present a shallow CMF index.

In Paper \citetalias{oneill2021}, a correlation between high mass surface density of cores and shallow CMF index was discussed. The results from the CMZ seem to follow the same pattern, if we consider the mass surface density of cores (see Table \ref{tab:core_data}). The region with the lowest core mass surface densities, i.e., the Brick, has the steepest CMF index and the region with highest core mass surface density, Sgr~B2, has the shallowest. In Figure \ref{fig:msd_vs_slope}, the index of the CMF is plotted against the mean and median mass surface densities of the cores for the six core populations studied in this paper series. With the exception of the IRDC sample from Paper \citetalias{liu2018}, there is a discernible trend towards shallower CMF slopes in regions with higher mass surface density cores.

Both these observations, that the CMF becomes shallower due to evolution and high mass surface density, can be explained if the cores accrete gas from the surrounding clump. According to the core accretion model of \cite{mckee2003}, prestellar and protostellar cores are expected to interact with gas from the surroundings with a rate that depends on the clump mass surface density, $\dot{M}_{\rm acc} \propto \Sigma_{\rm cl}^{3/4}$, although, as discussed in Section~\ref{S:intro}, it is unclear if this gas would become gravitationally bound to the core. This would mean that all cores become more massive as they evolve, but the effect is most noticeable in high-density regions. The accretion rate is also expected to be higher for more massive cores. The fact that we do not see any flattening of the CMF in the G286 protocluster can then be explained by the region's low mass surface density compared to the regions studied in this paper. The accretion may be too slow to make any difference in the CMF slope in the time it takes for star formation to commence.

However, an important caveat needs to be noted. If the protostellar cores in the dense regions are accreting more rapidly and thus more luminous, then they may be systematically warmer. This may lead to overestimation of core masses in these regions and derivation of a CMF index that is artificially shallow. The best way to resolve this possibility is to obtain more accurate core mass estimates, either via dynamical means, or via dust temperature measurements for each core.

\subsection{Mass Segregation and Clustering}

The mass segregation was found to be different between the three regions. While the Brick lacked any signs of mass segregation, Sgr~C showed indications of mass-segregation up to $\Lambda_{MSR}\sim 2$. The mass segregation in Sgr~B2-DS is of a similar scale, but more statistically significant. This means that a correlation between mass segregation and evolutionary stage can be seen. 
This fits in well with the cores accreting gas from the clump, which was discussed in Section \ref{dis:evolutionary_stage}. The cores that are located in the densest parts of the cloud are expected to accrete gas from the clump at a higher rate, causing them to grow more massive. Thus the most massive cores should be localized in the densest region rather than being randomly distributed in the cloud.

It is important to note that the mass segregation of $\sim 2$, although significant, is low compared to some other regions from the literature. For example, \cite{plunkett2018} found a mass segregation of $\sim 3.7$ in the star forming region Serpens South. The cores studied were all under 1 \msol, with the strongest mass segregation affecting cores above 0.04 \msol. Furthermore, \cite{dib2019} reported mass segregations of 3.8 an 8.8 in  the nearby regions Aquila and Corona Australis, and 3.5 in the W43 complex. Corresponding  segregated core masses were 6.2-19.7 \msol, 0.5-1.3 \msol~and 16-102 \msol.

\cite{dib2019} found a correlation between the $Q$ parameter and the star formation rate, finding that regions with a higher star formation rate tended to be more centrally concentrated (i.e., higher $Q$ values). 
This seems to be consistent with our results, since the Brick had a lower $Q$ value than Sgr~C, which is known to be forming stars. The northern part of the Sgr~B2-DS map had a $Q$ value similar to Sgr~C. As mentioned in Section \ref{res:spatial}, the low $Q$ value for the entire Sgr~B2-DS map is likely biased due to the shape of the map.

We can also compare our results to Paper \citetalias{oneill2021}, which calculated $Q$ values for the clumps in their sample with the most detected cores. These clumps were all classified as protostellar, and had $Q$ parameters in the range $0.67-0.82$. Furthermore, \cite{moser2020} calculated the $Q$ value for 35 protostellar cores in an IRDC, obtaining a value of $0.67$. These results are similar to our values for Sgr~C (0.71) and the limited Sgr~B2 map (0.67), but higher than the Brick value (0.52).

\cite{wu2017} performed simulations of star cluster formation in colliding and non-colliding clouds. In the non-colliding case, they found that $Q$ quickly stabilized at very low values ($\sim 0.2$). Our results are more consistent with the colliding case, for which the $Q$ value stabilizes at $\sim 0.6$. However, the simulations by \citet{wu2017} do not support a monotonic increase of $Q$ with evolutionary stage. In the colliding case, $Q$ first grows towards a peak above $0.8$, to then drop to $\sim 0.3$ before growing towards $\sim 0.6$ again. Note that these simulations are simplified and do not include feedback from the forming stars. Nevertheless, they indicate that the relationship between $Q$ and evolutionary stage could be complex.

When interpreting the mass segregation and $Q$ parameter for Sgr~B2-DS, it is important to keep in mind that the ALMA image shows the outskirts of a larger cloud complex, Sgr~B2. It might not be directly comparable to the Brick and Sgr~C, where the ALMA map shows the main cloud. The results should therefore be treated with caution.

\subsection{Implications for Star Formation Theories}

Top-heavy IMFs have been observed in the Galactic center, e.g., by \citet{lu2013} in the Nuclear Star Cluster
and \cite{hosek2019} in the Arches cluster in the CMZ. They found indices of $\alpha=0.7$ and $0.8$ respectively. The top-heavy CMFs in Sgr~B2-DS and Sgr~C could be consistent with a core accretion model with constant star formation efficiency, if the emerging IMF is also top-heavy. 

It is difficult to explain how a top-heavy CMF could turn into a canonical IMF. The issue is discussed in \citet{pouteau2022}, where they propose different shapes of the emerging IMF based on their observed CMF and various fragmentation scenarios and star formation efficiencies. Their CMF has a power law index of $0.95\pm0.04$, which is similar to our Sgr~C results with power law index $0.99\pm0.06$ 
(although derived by a different fitting method). In order to obtain a Salpeter-like IMF through fragmentation, they need to assume a number of fragments per core given by $N_{\rm frag}(M) \propto M^{0.4}$. This can be compared to the number of fragments for thermal Jeans fragmentation, which is given by $N_{\rm frag}\propto M$, under the assumption of constant density and temperature in all cores. So in order for Jeans fragmentation to explain the difference between CMF and IMF, the Jeans mass would need to be substantially higher in massive cores. This could be the case if the more massive cores have much higher gas temperatures. We note that the constant density assumption may not hold for the cores in this work, as the densities of cores in Sgr~B2 and Sgr~C appear to increase with mass (see Figure \ref{fig:mass-radius-relation}). However, this would make the Jeans mass smaller in more massive cores, and an even stronger temperature dependence would be needed to reach $N_{\rm frag}(M) \propto M^{0.4}$. Most other scenarios discussed by \citep{pouteau2022} either lead to an IMF that is even shallower than the CMF, or produce an IMF that is much steeper than the Salpeter slope.

The shallow CMFs observed in Sgr~C and Sgr~B2 could instead be consistent with a core accretion scenario in which the prestellar core population is influenced by environment, e.g., a bottom-up coagulation scenario for the cores (see Section~\ref{S:intro}). The CMF appears similar in shape to some estimates of the IMF in this region, which could imply a relatively constant value of $\epsilon_{\rm core}$ that is independent of core mass.

It remains to be seen if Competitive Accretion scenarios under realistic Galactic Center conditions can produce a top-heavy IMF \citep[see, e.g.,][]{2022MNRAS.515.4929G}. Given the uncertainties in modeling star formation, including properly accounting for the role of dynamically strong magnetic fields and protostellar feedback, the results of such simulations need to be treated with caution. More direct tests of predictions of Competitive Accretion can involve comparisons of spatial clustering and mass segregation metrics, and direct searches for clustered populations of lower-mass protostars around massive sources \citep[see, e.g.,][]{2024A&A...682A...2C}.

\subsection{Caveats}
\label{dis:caveats}

There are a few caveats that one needs to be aware of. Firstly, as mentioned, variations in core temperature may cause systematic errors in the mass estimation. As discussed in Section~\ref{met:mass_estimation}, cores that are warmer than we assume will have their masses overestimated. It is quite possible that the most massive cores in Sgr~B2 and Sgr~C host massive protostars, and thus have higher dust temperatures than the less massive, prestellar and protostellar cores. This would lead to an overestimation of the masses of the most massive cores specifically, causing the core mass function we derive to be biased towards shallower slopes. Recently, \cite{jeff2024} identified nine hot cores in Sgr~B2-DS, and estimated gas temperatures in them between 200 and 400 K. The positions of the cores all match with cores that are among the 20 most massive in our sample. \cite{jeff2024} argue that their cores are dense enough for the temperatures of the gas and dust to be well coupled. If this is the case, it could significantly affect the masses of the most massive cores in our Sgr~B2-DS CMF. However, since the affected cores are so few in number, assigning them the estimated gas temperatures or removing them from our sample does not have a significant effect on the CMF power law index. When doing so the MLE-derived $\alpha$ increases by $\sim0.02$, which is less than the standard error. Nevertheless, to make sure that the difference in CMF between the regions cannot only be attributed to temperature differences, better temperature estimates of the CMZ cores are needed.

More generally, the reliability of CMF measurements in distant regions, such as the Galactic Center, needs to be improved with a focus on samples selected to be prestellar cores \citep[e.g., selected to be highly deuterated, e.g.,][]{2013ApJ...779...96T,2017ApJ...834..193K,2021MNRAS.502.1104H}, development of higher-resolution MIR extinction mapping methods \citep[e.g.,][]{2012ApJ...754....5B}, more accurate temperature estimates at the core scale, and development and use of dynamical mass measurements of the cores \citep[][]{2020ApJ...894...87C}.

There is a possibility of free-free emission contributing to the mm fluxes, leading to an overestimation of core masses. To assess the importance of this, we checked for known ultra compact HII (UCHII) regions in the three clouds. \citet{lu2019} searched for UCHII regions in several Galactic center clouds, including the Brick and Sgr~C. They found no such regions in the Brick. In Sgr~C, four UCHII regions were detected, but their emission was found to be much weaker than the dust emission of the cores identified in the paper. The proportion of 1.3~mm flux density attributed to free-free emission was of the order of 1 \% for all affected cores in their sample. UCHII regions in Sgr~B2 were studied by \citet{meng2022}. They identified three such regions located within our Sgr~B2-DS map, two of which correspond to sources outside the ALMA primary beam FWHM, and a third which does not have a corresponding source identified by dendrogram. We therefore do not expect free-free contamination to be a major source of uncertainty in our core mass estimates.

Another notable caveat is that the resolution of our ALMA data is limited, with the beam corresponding to a physical scale of $\sim 0.02-0.04$ pc depending on the region.  It is possible that the larger cores, with diameters of a few beams, would appear fragmented if imaged with higher resolution (exemplified for the Brick in \cite{walker2021}, and for Sgr~C in \citet{lu2020}). The effect of resolution on the CMF was investigated in Paper \citetalias{cheng2018}, where a lower resolution was found to give a slightly shallower true CMF. The same tendency can be seen when smoothing the images of Sgr~C and Sgr~B2-DS to Brick resolution, as shown in Section \ref{sec:resolution}. We note that the analysis at uniform resolution leads to even greater differences in the CMFs of the Brick and the other two regions.


\section{Conclusions}
\label{S:conclusion}

We have measured the core mass function in three different clouds in the Central Molecular Zone, a region known to harbor extreme physical conditions compared to the local ISM. The regions are different from each other in that the Brick is mostly quiescent, while Sgr~C and Sgr~B2 are active star formation sites. A total of 711 cores were identified using the dendrogram algorithm in ALMA band 6 ($\sim 1\:$mm) continuum images. 

Flux correction and number correction was performed on the core samples, using a new method that takes core size into account. We also used this method to reanalyze the core samples from Paper \citetalias{cheng2018}, \citetalias{liu2018} and \citetalias{oneill2021}. The new correction method increased the number of high-mass cores compared to the previous method, but the difference in the derived CMF power law indices was modest. 

We fitted power laws to the CMFs above 2 \msol, using both the weighted least squares fit and a maximum likelihood estimator. We report the MLE parameters as our final results. For the Brick, an index of $1.28\pm0.09$ was found above 2~\msol, consistent with the Salpeter value. The CMFs of 
Sgr~C and Sgr~B2 were found to be relatively top-heavy, with indices of $0.99\pm0.06$ and $0.70\pm0.03$, respectively. The CMF of Sgr~B2-DS is significantly different from the other two according to the A-D test. If the maps are smoothed to the same resolution, Sgr~C and Sgr~B2-DS become more similar while the difference in power law index between the Brick and the other two becomes more pronounced. Fitting from the bin centered at 3.9 \msol, which is the lowest bin populated for all smoothed CMFs, we obtain $\alpha=1.36\pm0.12$ for the Brick, $\alpha=0.66\pm 0.06$ for Sgr~C and $\alpha=0.62\pm 0.04$ for Sgr~B2-DS. 

We analyzed the spatial distribution and mass segregation of cores by calculating the $Q$ parameter and the mass segregation parameter $\Lambda_{\rm MSR}$. The $Q$ parameter was notably smaller for the Brick ($Q=0.52$) than the other two regions ($Q=0.71$ for Sgr~C and $Q=0.67$ for the densest part of Sgr~B2-DS). The values indicate that all the regions are substructured rather than radially concentrated, but that the Brick has the highest degree of substructure. The Brick also showed no evidence for mass segregation. Sgr~C showed a mass segregation of $\sim 2$ for the 8 most massive cores, but the difference from 1 was just below statistical significance. Sgr~B2 was seen to be significantly mass segregated at a level of $\Lambda_{\rm MSR}\sim 2$ for the 5-11 most massive cores, and a lower $\Lambda_{\rm MSR}$ value for the 12-14 most massive cores. The values of $\Lambda_{\rm MSR}$ suggest that the mass segregation is related to the evolutionary stage of the clumps.

To the extent that the CMF results are not affected by systematic temperature variations, they imply that statistically significant variations of CMF (or more accurately core mm luminosity functions) can occur in different environments and at different evolutionary stages of star cluster formation. Scenarios based on Core Accretion and Competitive Accretion can be tested against the data we have presented. As we have discussed, Core Accretion scenarios in which prestellar cores grow to higher masses in dense regions or Competitive Accretion models involving significant clump-fed mass accretion during the protostellar phase may be consistent with the observed trends, but development of more detailed and realistic models is needed. 

Higher resolution and more sensitive mm continuum observations of these regions are needed to probe to lower masses, in particular to search for the peak of the CMF, which may be a more robust metric against which to test star formation models. At the same time, the reliability of CMF measurements in these distant regions needs to be improved with a focus on samples selected to be prestellar cores, development of higher-resolution MIR extinction mapping methods of such cores, more accurate temperature estimates at the core scale, and development and use of core dynamical mass measurements.


\begin{acknowledgments}
We thank Theo O'Neill for helpful discussions. JCT acknowledges funding from ERC Advance Grant MSTAR 788829. GC acknowledges support from the Swedish Research Council (VR Grant; Project: 2021-05589) and the European Southern Observatory. This paper makes use of the following ALMA data: ADS/JAO.ALMA\#2012.1.00133.S, ADS/JAO.ALMA\#2016.1.00243.S, \newline  ADS/JAO.ALMA\#2017.1.00114.S.  ALMA is a partnership of ESO (representing its member states), NSF (USA) and NINS (Japan), together with NRC (Canada), MOST and ASIAA (Taiwan), and KASI (Republic of Korea), in cooperation with the Republic of Chile. The Joint ALMA Observatory is operated by ESO, AUI/NRAO and NAOJ.
\end{acknowledgments}

\vspace{5mm}
\facilities{ALMA, CSO (Bolocam)}

\software{astropy \citep{astropy2013,astropy2018,astropy2022},   
          astrodendro (\url{http://www.dendrograms.org/}), , scipy \citep{scipy2020}, CASA \citep{casa2022}. }

\appendix
\section{Convergence of recovery curves}
\label{app:convergence}
The following plots show how the flux recovery, number recovery and radius recovery fractions described in \S\ref{met:recovery} change with each iteration. Sgr~B2-DS was chosen as a representative case.

\begin{figure*}
\gridline{
    \includegraphics[width=0.45\linewidth]{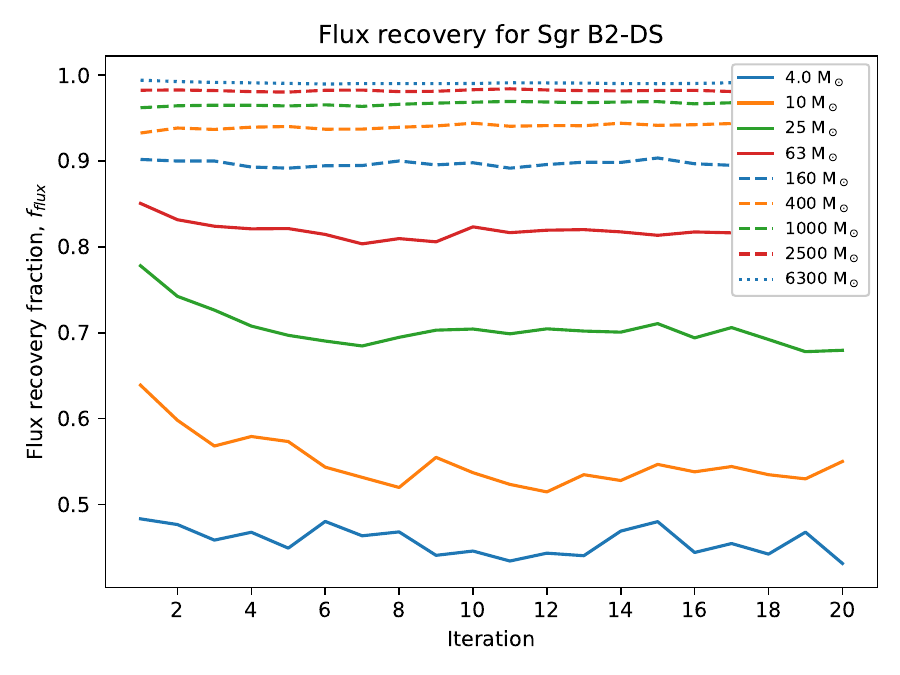}
    \includegraphics[width=0.45\linewidth]{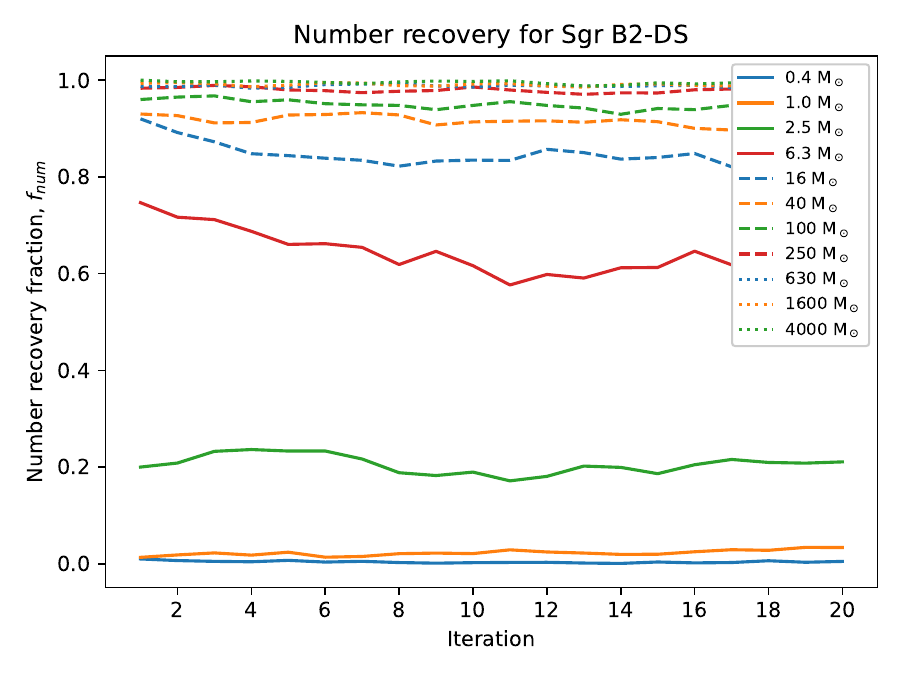}
    }
\gridline{
    \includegraphics[width=0.45\linewidth]{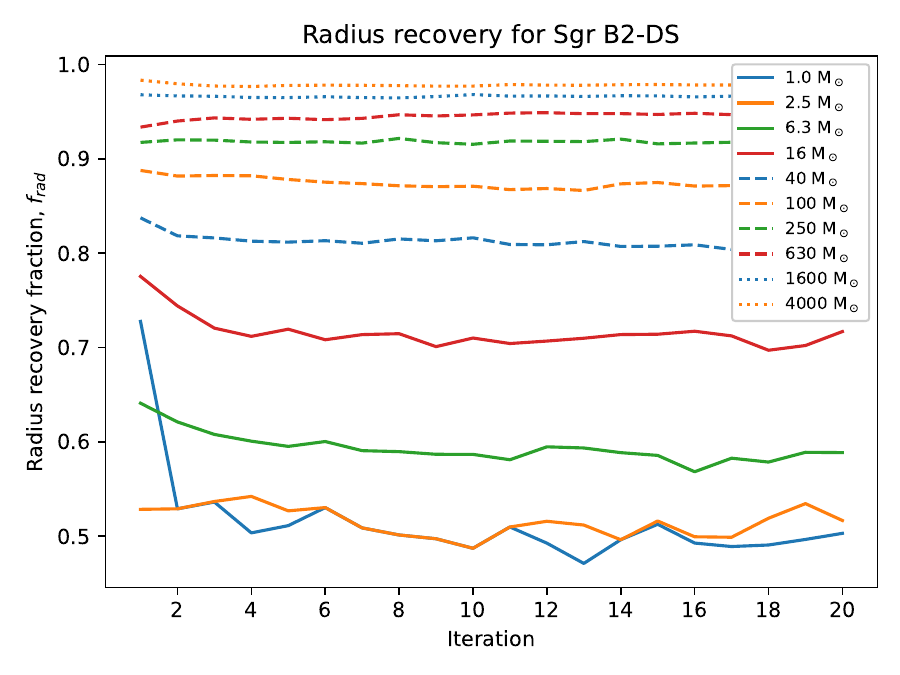}
    }
\caption{Convergence plots for flux recovery, number recovery and radius recovery fractions for Sgr~B2-DS. Each line represents one mass bin, plotting its recovery fraction versus iteration number. For readability, only every other mass bin is shown. 
We see that the high-mass bins are stable across all iterations, while some lower mass bins exhibit variation in early iterations, but stabilize later. }
\label{fig:convergence}
\end{figure*}

\bibliography{references}{}
\bibliographystyle{aasjournal}

\end{document}